# Understanding the Dynamics and Optimizing the Performance of Chemostat Selection Experiments

Aryeh Wides[1], and Ron Milo[1]*
[1] Department of Plant and Environmental Sciences, Weizmann Institute of Science, Rehovot 7610001, Israel
*Correspondence: ron.milo@weizmann.ac.il

## ABSTRACT

A chemostat enables long-term, continuous, exponential-phase growth in an environment limited as prescribed by the researcher. It is thus a potent tool for laboratory evolution - selecting for strains with desired phenotypes. However, despite the apparently simple design governed by a limited set of rules, analysis of chemostat dynamics shows that they display counter-intuitive properties. For example, the concentration of limiting substrate in the chemostat is independent of the concentration in the influx and only dependent on the dilution rate and the strain parameters. Moreover, choosing optimal operational parameters (dilution rate, volume size, influx substrate concentrations) can be challenging. There are conflicting requirements in the experimental design, such as a need for relatively fast growth conditions for mutation accumulation on the one hand versus slow dilution to confer a large relative fitness advantage to mutants so that they take over the population quickly on the other.

In this study, we provide analytic and computational tools to help understand and predict chemostat dynamics, and choose suitable operational parameters. We refer to five stages of the process: (A) parameter choice and setup, (B) basic "steady state" growth, (C) mutation occurrence, (D) single takeover and (E) successive takeovers. We present a qualitative and quantitative framework to answer the questions confronted in each of these stages. We provide a set of simulations which support the quantitative results, and a graphical user interface to give a hands-on opportunity to experience and visualize the analytic results. We substantiate the analysis by revisiting published selection studies.

In terms of experiment design, we detail conditions that produce ineffectual selection regimes, and find that when these parameter regimes are avoided, the selection time is relatively robust, and usually varies by less than an order of magnitude. Finally, we suggest rules of thumb to help ensure that the chosen chemostat parameters lead to effective selection and minimize the duration of the selection process.

The paper is arranged for convenient use, in the format of questions followed by short answers, complemented with full derivations in the appendices.

# Table of contents

# I. A guide for the chemostat perplexed

This manuscript contains a lot of information about chemostats on different levels. It can be used for a range of purposes, from the relatively simple and practical to the quantitatively-demanding technical: The manuscript can hopefully be used as an aid for the setup of chemostat experiments, and as a tool for an informed choice of experimental parameters, and ultimately to advance a deeper understanding of chemostat dynamics. The document is therefore built in a layered fashion, to suit the needs of different readers, as follows:

I. **A "guide for the chemostat perplexed"** - Here we start off with a quick and short guide, describing when and why chemostats can be useful, together with basic practical steps one should take when setting up a chemostat selection experiment.
II. **Non-intuitive insights -** Next, we bring an incomplete list of what we think are non-trivial and non-intuitive findings about chemostat operation in general and selection experiments in specific that you will learn from reading this manuscript. Beyond knowledge, we hope you will draw intellectual fun as we did from understanding these. The list is ordered by the chapters of the paper, and includes pointers to the relevant sections
III. **Concise Q&A** - The main body of the paper consists of five chapters in the format of questions-and-answers, (A) parameter choice and setup, (B) basic "steady state" growth, (C) mutation occurrence, (D) single takeover and (E) successive takeovers. The questions are listed in the Table of Contents, and the answers are on the order of half a page, bringing the main findings, without the mathematical derivations.
IV. **Comprehensive Q&A analysis** - Finally, a collection of appendixes is presented for the advanced reader. These are geared for those interested in the full mathematical derivations leading to the findings in the concise Q&A section.

**When and why should one use a chemostat**

A diagram of a chemostat is given in Figure 1. The design is simple: a well-mixed growth vessel contains your culture while media flows in at the same rate that media with cells flows out. This clever design ensures that once steady state will be established, cells will divide at exactly the same rate that you dilute the culture with fresh media (the dilution rate D, see below). As such, you can control the growth rate of your culture over a very wide range by setting the dilution rate, from near stagnation to close to the maximal growth rate of the cells. As we will see below, setting the dilution rate will also set the concentration of the limiting resource in the growth vessel. Cultures can be kept in continuous exponential growth for weeks or months, allowing for continuous application of relatively uniform selection pressure. A chemostat enables careful control of the cell growth rate, of the chemical environment, and of the selective pressures. By way of contrast, in propagation under manual serial dilutions experiments which are quite popular and useful, the limiting factor cannot be directly controlled and kept constant.

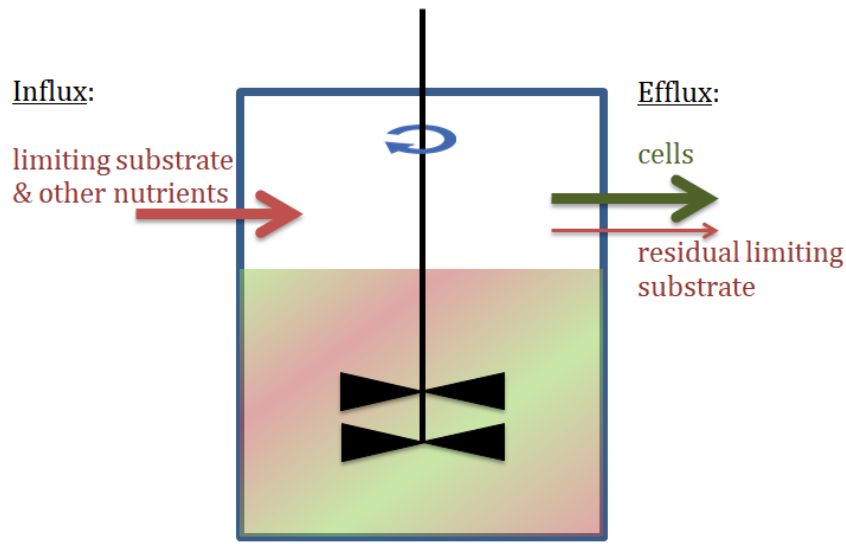

**Fig. 1:** Basic diagram of a chemostat. Influx of sterile media with nutrients and limiting substrate. Efflux of waste cells and media.

In methods of growth such as plating, serial dilution, batch growth, etc. a starting environment with limiting conditions will be depleted quickly, not allowing long term continuous growth; and starting with an abundant nutrient source does not allow growth under a high selective pressure. In both, the growth and substrate conditions do not stay constant, but rather start off at a relatively quick rate and slow as the resources diminish and saturation is reached. In step wise serial dilution this non-ideal process is similarly happening, only repeatedly. A chemostat enables careful control of the cell growth rate, of the chemical environment, and of the selective pressures.

Other growth apparatus, such as the turbidostat, auxostat, morbidostat, etc. are more complex experimentally (since they require some form of sensory feedback, e.g. based on online OD reading, that the chemostat does not require), and are not essential for many applications. The added complexity can, on the one hand, enable specific capabilities that are more difficult to achieve with the simpler chemostat, but on the other hand can generate complications that do not arise in the more robust chemostat. For example, the turbidostat enables growth at the maximal growth rate without the risk of washout, but on the other hand generates a risk of biofilms sitting on the OD reader and thus obstructing the control, and of other turbidity factors that can complicate the experiments (see section B8 for further elaboration).

Suffice for now is to appreciate that in short, a chemostat is the simpler approach and often suffices for most continuous growth experimental needs.

### Basic practical steps to setup a chemostat experiment
In the chemostat there are three main controllable parameters: the chemostat **volume**, the **dilution rate** and the **influx concentration** of limiting substrate.
Interestingly, the influx concentration will affect the steady state OD and therefore population size, but not the limiting substrate concentration within the chemostat (see sections A3-A4).

There are two relevant processes that occur repeatedly: mutation, and strain takeover (see sections B1-B2). The former takes a substantially shorter time when the population is big enough (see chapter C) and is thus relevant for the choice of chemostat volume and the limiting substrate influx concentration. The occurrence of strain turnover in a reasonable amount of time, which will lead to the selection of a desired strain, is affected by the choice of dilution rate (see chapter D).

The following four steps should be taken when setting up an experiment:
1. **Measure the growth parameters of the primary reference strain:**
   **Maximal growth rate**, **yield** (the mass of cells formed per unit mass of substrate consumed) and **K Monod** (the limiting substrate concentration for which the cells grow at half of the maximal growth rate, as formulated by Monod [1]). These are needed for the ballpark choice of operational parameters, so they do not have to be very exact, e.g. +-10% would be fine.

2. **Choose the main operation parameters for the chemostat:**
   a. If technically possible, the chemostat volume and limiting substrate concentration should ideally be chosen so that the population size is sufficiently large so that mutant generation rate is not limiting the temporal dynamics. Approximately $10^9$ cells for *E.coli* should suffice for single mutations (see section A4 for the equation connecting the parameters of the chemostat to the population size; and section C2 for why this is important).
   b. The influx concentration for other nutrients and substrates should be chosen so that they are non-limiting. This usually happens naturally as for the influx media one typically starts from a standard well balanced media and decreases the limiting substrate concentration and thus the other nutrients will be in excess and not limiting.
   c. While the equation depicting a chemostat is continuous, in reality it is implemented via small influx (and efflux) pulses of media which effectively achieve the same effect (this is termed quasi-continuous, see section A1). If not otherwise needed, the influx pulse volume should be relatively small, e.g. less than 1% of the chemostat volume, so that the growth will be quasi-continuous (if a larger influx is required for some reason - see section B4-B5 for a quantification of the effect).

3. **Choose dilution rate, dependent on the desired mutant improvement:**
   a. The desired improvement of a mutant strain can be either a **higher maximal growth rate**, an **ability** to grow without limiting substrate, a **higher affinity** to the limiting substrate or **any combination** of the above. Different dilution rates select for different improvements (see sections D3-D7). Note, different reasons for achieving a faster growth rate at a low substrate levels (such as by having better transporters, a more efficient utilization of the nutrients, or by any other improvement) can be categorized as an overall "higher affinity to the limiting substrate".
   b. A dilution rate which equals half the maximal growth rate is often a robust choice. The dilution rate should not be too fast nor too slow, so as to promote multiple successive takeovers on the way to the desired mutant strain (see section C3 for why successive takeovers are important, and section D5 for why extreme dilution rates are problematic).

4. **Expected operation time**:
    a. A successful mutation and population takeover is expected to take on the order of weeks, as opposed to days; more so when the desired strain requires multiple successive takeovers (see chapter D).
    b. For large inoculations there is a risk of "overshoot/undershoot" where the time to reach steady state is prolonged by several chemostat turnover times. This delay can be solved (see section A7).

# II. Non-intuitive insights about operation and selection in chemostats

Following is an incomplete sample list of things that we found non-intuitive about chemostat operation in general and selection experiments in specific that you will learn from reading this manuscript. The list is ordered by the chapters of the paper, and includes pointers to the relevant sections in parentheses.

A) PARAMETER CHOICE and SETUP
- The steady state concentration of the limiting substrate in the chemostat is independent of the influx concentration (A3). The influx concentration will affect the cell concentration and thus the steady state OD (A4).
- Even though the limiting substrate concentration in the chemostat is usually very low (A3), and is maintained by discrete highly concentrated influx pulses, in practice the temporal variation in the concentration within the chemostat is small (a few percent or less) and can thus be viewed as quasi-steady state. (B3-B4)
- The time it takes for the cell density (OD) to converge to a steady-state value (overshoot/undershoot) will often be long (multiple chemostat turnovers), especially when the initial inoculum is large. But, the time can be minimized with proper parameter choice. (A6-A7)
- Evaporation, in reasonable amounts, does not change the chemostat dynamics. (A8)

B) STEADY STATE
- A chemostat might appear to be in steady state, but mutant strain takeovers can occur continuously, even though they are not detectable by monitoring macro scale parameters like OD or product concentrations (B1-B2).
- The limiting substrate is usually at such low concentrations that it is undetectable. As a result, the concentration of the limiting substrate can vary greatly over time (percentage-wise) as different strains takeover the population, even if resulting changes in OD are too small to detect (B2).
- A "pulsed" chemostat (with very large influx pulses) has a substantially lower selective capacity than a standard quasi-continuous chemostat, for a mutant strain with increased fitness in limiting conditions (B4).
- By abruptly lowering the influx limiting substrate concentration it is possible to temporarily subject the cells to relatively harsher conditions, until the chemostat stabilizes back to the steady state (on the time order of the dilution rate D)(B6).

C) MUTATION
- Some types of mutant strains will appear rapidly:
    - If there is a SNP that can increase fitness it should appear in the population after only few chemostat doublings, for characteristically large chemostats (e.g. $10^{11}$ E. coli cells) (C1-C2).
    - A strain that requires two specific SNPs where only their combination gives a fitness advantage (whereas each one separately is neutral), is likely to appear only if the target size (the number of different SNP locations that give rise to an advantageous mutation) for each SNP is very large (C3-C4).
- Other types of mutant strains (e.g. two SNPs with a small target size, more SNPs or in smaller chemostats) are highly unlikely to appear (C5).
    - These other mutations are expected only through successive sweeps of mutants with a fitness advantage. We only expect multiple mutants to arise if each mutation is independently beneficial, and not in cases where the mutations are individually neutral but together

- advantageous. Successive takeovers are the only reliable way for evolution to proceed in a chemostat (C5).
- The seemingly extreme scenario where we require every possible single SNP to co-exist at least once in the chemostat is actually quite likely. A large chemostat is very likely to reach this state (C6).
- For a large chemostat we expect the time until an advantageous mutation occurs ($T_{mutant}$) to be on the order of the chemostat turnover time. Note, this is usually substantially shorter than the time for an advantageous strain to take over the chemostat population ($T_{takeover}$, see following). This is not necessarily so in a small chemostat (C).
- The above points are expected to be the same across different asexually reproductive species (*E. coli, S. cerevisiae,* etc.). (C7)
- Furthermore, the time until mutation appearance ($T_{mutant}$) is independent of genome size, but dependent on per-BP mutation rate (C7).
- For characteristically large chemostats, a hyper-mutating strain does not give enough of an advantage to warrant use (C8). Also, it does not have enough of a selective advantage to be expected to always appear through random mutation and take over the chemostat (C8).

D) SINGLE TAKEOVER
- The takeover time is predictable given the relevant strain parameters (D3-D6)
- Different dilution rates selectively favor different mutant strains to take over the chemostat population, if such a strain exists (D1-D7). For example:
    - A fast dilution rate creates a selection pressure for a mutant strain with a raised maximal growth rate (D4);
    - A mid-range dilution rate creates a selection pressure for a mutant strain with a higher affinity to the limiting substrate (D5);
    - A slow dilution rate creates a selection pressure for a mutant strain which can grow in media with no limiting substrate (presumably by consuming a different substrate present in the media) (D6);
- The time for takeover of a superior mutant ($T_{takeover}$) will be quite constant across a range of operation parameters. For characteristic operation values the take over time is on the order of days to weeks.

E) SUCCESIVE TAKEOVERS
- When the conditions are right (a large enough population, and multiple targets in the genome for simple advantageous mutations) multiple strains are expected to successively takeover the population, and to do so in a relatively timed and paced manner. This was also observed by Egli et al [2] . The timing depends on the type of mutations (E1-E3).
- In a takeover succession, even if the selective improvement of each of the strains stays constant (e.g. each new strain is better than the previous strain by a constant factor) – the takeover rate does **not** stay constant, but rather diminishes from strain to strain (E1-E4).
- There are cases where successive takeovers occur so rapidly that it is very difficult to differentiate between strains, even when examining allele frequency. Thus, a lineage of multiple takeovers of consecutive strains might appear as the takeover of a single strain with a cohort of mutations (E4). This finding can help explain experimental cases that have previously been only partially deciphered.

# List of parameters

Following are the parameters used in this essay (table 1). Throughout the text we present graphs and numerical "typical results" to complement analytic equation-based descriptions, to aid understanding. For this we define here a "default characteristic case" with assigned parameter values. These values are characteristic of a relatively standard lab setup. Specifically, they were based on, and used for, the selection experiments described in [3].

The parameters are illustrated in figures 1 and 2 below.

*Table 1) Parameters and default characteristic case values*

| Notation | Parameter | Default value | Units |
|---|---|---|---|
| Chemostat parameters (directly controllable) | | | |
| The first set of parameters consists of parameters that determine the macro-dynamics of the chemostat, and are measurable (potentially) and experimentally controllable on the global scale. | | | |
| $V_C$ | **Chemostat tank volume** | 0.5 | |
| $V_d$ | **Dilution pulse volume** - Most chemostats do not have a truly continuous influx of media, but rather are "pulsed". At every time interval $\tau_d$ a pulse of fresh media of volume $V_d$ is inserted into the tank, and a corresponding volume flows out. When $V_d$, $\tau_d$ are sufficiently small we can treat the system as continuous. For most chemostats, and for our test case, the influx comes as individual drops | $10^{-4}$ ($= 100\mu L$) | [L] |
| $\tau_d$ | **Dilution interval** - can vary greatly, depending on D | 7 [s] (for D = 0.07 h⁻¹) | [s] |
| $\phi$ | **Volumetric dilution ratio** - Ratio of media remaining after dilution step $\phi = 1 - \frac{V_d}{V_C}$. For most chemostats dilution process can be treated as continuous, $\phi \to 1$. | 99.98% | *Unitless* |
| $D$ | **Dilution parameter** - Fraction of the chemostat volume replaced per hour by inflow from the reservoir. $D = \frac{V_d}{V_C} \cdot \frac{1}{\tau_d} = \frac{1-\phi}{\tau_d}$. D can vary from 0 (no dilution) to the maximal growth rate of cells (above which the chemostat will be flushed out of cells). | 0.07 (when not used as a parameter) | $\left[\frac{1}{h}\right]$ |
| $S_d$ | **Limiting substrate influx concentration** | 0.1 | $\left[\frac{g}{L}\right]$ |
| Strain parameters (inherent to strain. Sub-index i denotes strain) | | | |
| The additional set of parameters consists of properties of the various cell strains. **Subindex $i = 0$** denotes original reference strain, while **subindex $i \geq 1$** denotes the mutant strain after i successive takeovers. Note, the units used for all concentrations, following and above, are $\left[\frac{g}{L}\right]$ so as to be congruent with each other. | | | |
| $\mu_i^{max}$ | **Maximal growth rate** - cell growth rate when limiting substrate is abundant | 0.17 | $\left[\frac{1}{h}\right]$ |
| $\mu_i^{min}$ | **Minimal growth rate** – cell growth rate when no limiting substrate is present | 0 | $\left[\frac{1}{h}\right]$ |
| $K_i$ | **Parameter for Monod relation for limiting substrate** – the limiting substrate concentration for which the cells grow at half of the maximal growth rate | $10^{-3}$ | $\left[\frac{g}{L}\right]$ |
| $Y_i$ | **Yield** - the mass of cells or product formed per unit mass of substrate consumed, for strain i | 2 | *Unitless* |

| | Population-related chemostat parameters (indirectly controllable) | | |
|---|---|---|---|
| colspan=4 | The next set of parameters consists of parameters that are measurable (potentially) on the global scale, but are dependent on the cell strains in the population. These parameters are not directly controllable, but can be set using the chemostat parameters (above) and knowledge of the strain parameters (as will be shown in the body of this paper). | | |
| $S(t)$ | **Limiting substrate concentration in chemostat** - the free limiting substrate concentration, not including the limiting substrate that has been consumed by the cells. The exact value is time dependent. | - | |
| $S_i^{st}$ | **Steady state limiting substrate concentration when the chemostat is dominated by strain i** - right before an influx drop. Note, the difference between $S(t)$ and $S_i$ is that the first is a time dependent value and the second is a steady state constant value. | $7 \cdot 10^{-4}$ | $\left[\frac{g}{L}\right]$ |
| $X_i(t)$ | **Cell strain i dry weight (or concentration) in chemostat.** The exact value is time dependent. | - | |
| $X_i^{st}$ | **Steady state cell dry weight in the chemostat when the chemostat is dominated by strain i** | $\approx 0.2$ | |
| $N_{cells_i}$ | **Number of cells of strain i** | $2.5 \cdot 10^{11}$ | Cells |
| $OD_i$ | **Optical density of strain i.** As will be shown, to convert $X_i \leftrightarrow OD_i$ for *E. coli* use $1 [OD_{600}] = 0.4 \left[\frac{g\ CDW}{L}\right]$ | 0.5 | 1 cm path length |
| $\mu_i(t)$ | **Specific growth rate of strain i** – defined as $X_i \cdot \mu_i(t) = \frac{dX_i}{dt}$. The exact value is time dependent. | - | $\left[\frac{1}{h}\right]$ |
| colspan=4 | Species parameters – mutation | | |
| colspan=4 | The final set of parameters is inherent to the cell species, and affect the time until a mutant strain is expected to appear. They do not change during the experiment. | | |
| $G$ | **Genome size** | $5 \cdot 10^6$ *E. coli* $10^7$ *S. cerevisiae* | $[bp]$ |
| $\theta$ | **Target size** – the number of different SNP locations that give rise to an advantageous mutation | ~1-1000 | $[bp]$ |
| $R_{mut}$ | **Point mutation rate** | $10^{-10}$ | $\left[\frac{SNP}{bp * doubling}\right]$ |

**Other parameters in use**

- $T_{Steady\ State}$ – Time from setup inoculation until the chemostat reaches steady state for the reference strain. We define "reaches" as when the difference between the theoretical steady state concentrations (for OD, substrates, etc.) and actual concentrations due to the transient setup process - is indistinguishable from random noise fluctuations.
- $T_{mutant}$ – Time until the desired mutant strain appears in the chemostat.
- $T_{takeover}$ – Time from the first appearance of a mutant until it takes over the reference population, which we define as the strain reaching at least 10% of the chemostat population.

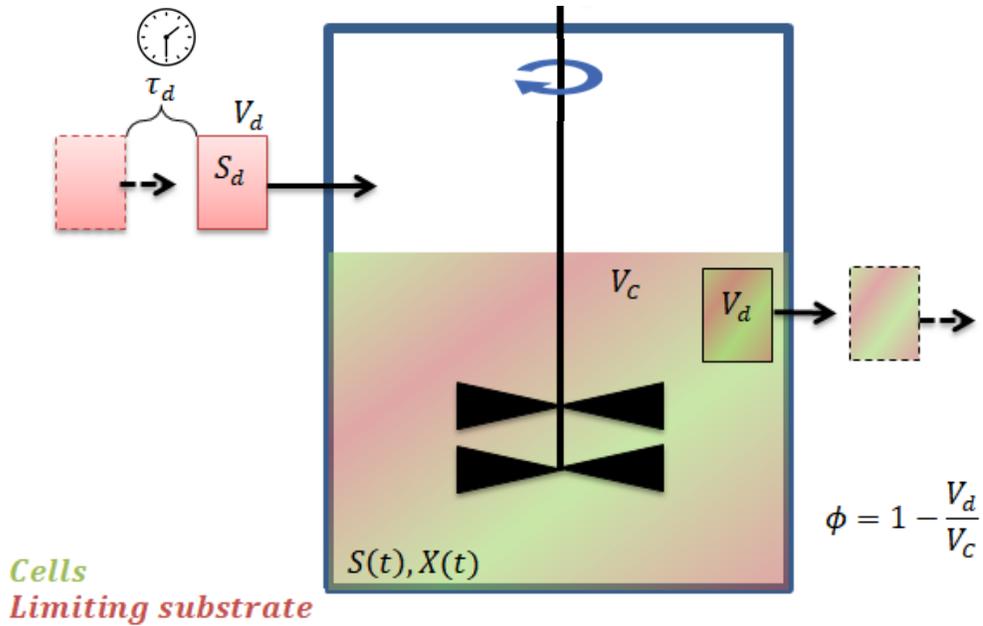

**Fig. 2a: Chemostat parameters:** The chemostat is of volume $V_C$, with a cell concentration of $X(t)$ and a limiting substrate concentration of $S(t)$. At every time step $\tau_d$ [h] a dilution influx of volume $V_d$ and limiting substrate concentration $S_d$ is added to the chemostat, while a corresponding outflux of the same volume is removed, with concentrations $X(t)$ and $S(t)$ as in the chemostat (in the figure, red representing limiting substrate and green representing cells). The ratio of media remaining after dilution is $\phi$.

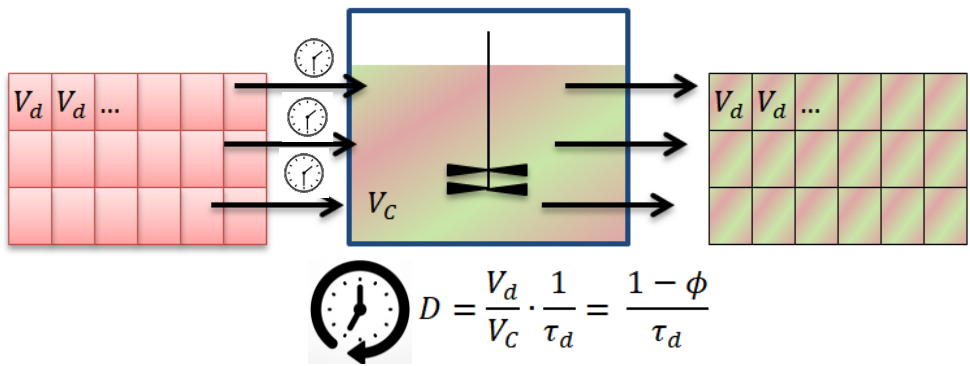

**Fig.2b:** D is the fraction of the chemostat volume replaced per hour by influx; in other words, 1/D corresponds to the amount of time until the cumulative volume of the influx (or equivalently efflux) equals the volume of the chemostat.

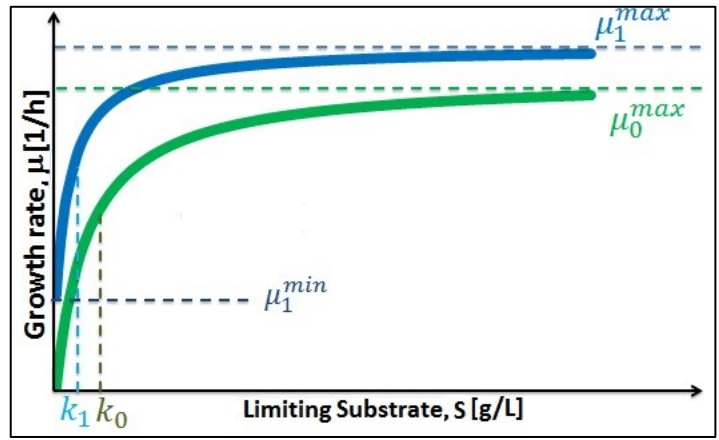

**Fig. 3: Strain parameters:** In green is the Monod curve of reference strain (growth rate $\mu_0$ as a function of limiting substrate S), with a maximal growth rate $\mu_0^{max}$, and a growth rate of $\frac{\mu_0^{max}}{2}$ for limiting substrate concentration equaling the Monod parameter $S = K_0$. In blue is the expanded Monod curve for mutant strain, $\mu_1 = f(S)$, with a maximal growth rate of $\mu_1^{max}$, Monod parameter $K_1$ and minimal growth rate of $\mu_1^{min}$. A value of $\mu_1^{min} > 0$ describes a case where the limiting substrate (S) is not the sole carbon source, and the mutant strain can grow without use of S.

## Default color scheme for graphs

To aid quick understanding of graphs, we shall use a coloring system, used unless stated otherwise. A green line will usually denote the reference strain while a blue line will denote the mutant strain. A red line will denote the limiting substrate concentration.

# Introduction

## Introduction to the chemostat

Novick and Szilard invented the chemostat in order to keep a bacterial population growing continuously at a fixed, user-defined growth rate for an extended period of time (hundreds of generations) [4]. The chemostat consists of a culture tank, an influx stream of nutrients in sterile medium, and an output stream of a corresponding amount of medium containing cells and waste. After a sufficient amount of time, the cell population and nutrient concentration in the vessel reach a steady state wherein the rate of cell growth is exactly balanced by the rate of efflux (the dilution rate). That is, the growth rate $\mu$ equals the dilution rate D at steady-state. Therefore, $\mu$ can be manipulated over orders of magnitude. By adjusting the dilution rate, the growth rate in the chemostat can be varied greatly, from near stagnation to the maximal growth rate of the cells [5].

Here we will describe and analyze the use of chemostats as an apparatus for attaining a desired strain through selection in a substrate limited environment. "Substrate limited" in the sense that not all nutrients are abundant enough for growth at the maximal growth rate, for instance, an environment where the carbon source concentration is very low, slowing down growth. Since the chemostat enables continuous growth in an environment of constant substrate-limited conditions, it enables selection for strains with improved growth in those conditions. More specifically, chemostats select for increased growth rate in limiting conditions (as we shall see further on). Hence, they can be used to select for a randomly appearing mutant strain out of a reference strain, where the mutant has a higher specific growth rate[3], [5], [6], [7].

For instance, suppose you want to select for *E. coli* cells that grow more quickly on xylose as their sole carbon source. To do so, you would try to cultivate *E. coli* in a growth environment where more efficient xylose metabolism increases their growth rate. To find such an environment, you would measure the growth rate as a function of xylose concentration (i.e. the Monod curve [1]) and cultivate cells in batch culture in the xylose concentration that gives half-maximal growth, for example. In such conditions we'd say that xylose is "growth-limiting," meaning that increasing the media xylose concentration would increase the growth rate of the culture. Intuitively, these conditions put the whole culture in competition over xylose - any cell that can access or metabolize xylose more efficiently will grow faster than the rest of the culture and, therefore, increase in proportion.

For another example, in [3] a cell strain of *E.coli*, which can grow independently of xylose, was evolved through selection from a reference strain that is xylose dependent. This was achieved in a xylose-limited chemostat (conditions where the desired mutant strain had a higher growth rate).

## Goals and statement of problem

How can we better understand processes in the chemostat in order to design reproducible selection experiments? In addition to having reproducible selection experiments, we would like improved strains to arise quickly. How can we choose selection parameters that increase the chance of improved strains arising in a reasonable amount of time?

In other words, given a chemostat with a limiting substrate (e.g. xylose) and a particular reference strain of cells – how long will it take for a particular improved mutant to take over the chemostat population? How can the result be controlled by regulating the parameters of the chemostat setup and depending on the parameters of the cell strains (ref. and mutant)?

In this study we offer tools and rules of thumb to optimize the design of selection experiments in two ways: The first, to help avoid pathological cases where the selection process is not expected to succeed, such as washout - which occurs when the dilution rate is too fast for the reference strain to match and the vessel is emptied of cells. The second, to asses and minimize the amount of time selection takes, by careful choice of controllable chemostat parameters, and depending on uncontrollable species, strain and population parameters. We refer to five stages of the process: (A) parameter choice and setup, (B) basic "steady state" growth, (C) mutation, (D) single takeover and (E) successive takeovers. When a mutant of increased fitness arises and grows to replace the original population, we call this a "takeover."

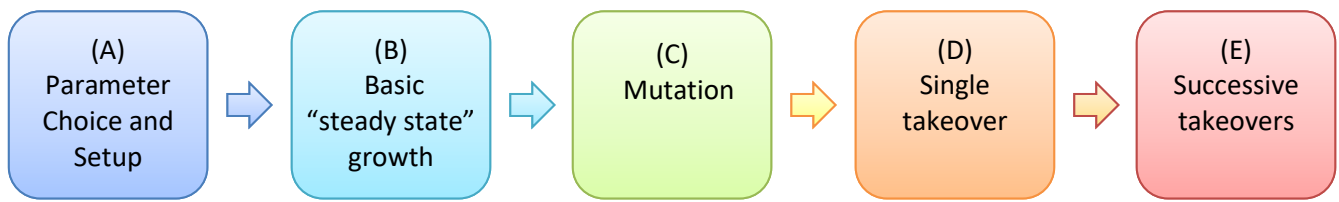

*Fig. 4:* Outline of the chapters in this paper, in reference to the stages of a chemostat selection experiment.

The chemostat dynamics are not straightforward, with contradicting processes occurring simultaneously, so that the choice of experimental parameters is often neither simple nor intuitive.

For example, when using a chemostat for selection, the device has at least two contradicting roles. The first is the creation of an appropriate mutant strain. The rate of appearance of new mutants is directly correlated with the number of doublings [8]; hence, constant growth in good conditions (abundant limiting substrate leading to fast growth) is preferable. The second role is the selecting for an improved mutant strain. Since limiting conditions are expected to improve the selective advantage of the mutant, growth in nutrient poor conditions (i.e. slow growth) is preferable for selection. Therefore, the considerations associated with generating and selecting for improved mutants are in tension with each other - there is a tradeoff between fast growth (quick generation of new mutations) and slow growth (higher relative advantage for improved mutant strains).

Another example is the choice of the growth conditions after a desired mutation succeeded, for the maximal selective advantage between the dominant reference strain and the new mutant strain: In some cases (such as an improved affinity to the limiting substrate), if growth conditions are too limiting - the growth rate of the mutant strain will be reduced to a point where the selective advantage is diminished. If the growth conditions are not limiting enough – the reference strain will be growing at a rate in which the mutant strain does not have a substantial selective advantage (see Fig. 5).

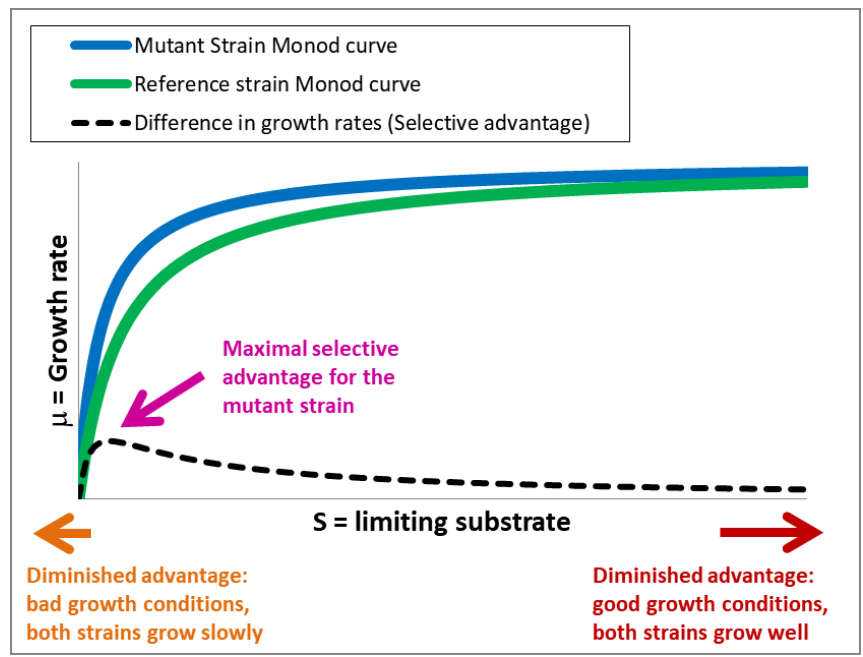

*Fig. 5: Green) Monod curve of reference strain. Blue) Monod curve for mutant strain. Black) difference in growth rates. Red & Orange) extremes in growth conditions (limiting substrate concentration) leading to a diminished selective advantage for the mutant strain, meaning longer selection time. Magenta) limiting substrate concentration for which selective advantage is maximal.*

# Previous work

The chemostat is a widely used apparatus, and there are tens of thousands of published papers involving chemostats. The majority of these studies are of an experimental nature, where the chemostats are used as tools, but not studied directly. Studies that include, at least in part, theoretical modeling of the chemostat takeover and selection process – usually focus on a certain aspect of the progression.

The primary studies that set the basic theoretical groundwork are of Novick & Szilard [4] which describes the bioreactor and coins the term "chemostat"; and Dykhuizen & Hartl [5] which discusses chemostat theory, including specifying the important parameters and their relationships, and analyzes mutation and selection.

There are experimental studies that include some theoretical discussions, such as [9] which primarily uses an experimental model to test the axiom that a "generalist" is less efficient than a "specialist", but also has a theoretical chapter, where the growth equations are studied to help predict the percentage generalist at equilibrium. Another example is [10] which experimentally shows an increasing affinity to the limiting substrate through selection but also has a short discussion concerning the connection to the growth equations. Similarly, [11]–[14], among many others, are primarily experimental but have some theoretical discussion.

Some mathematical studies model certain elements, special setups, or variations to the standard growth models in the chemostat. Examples of such studies include: [15] models competition for two limiting substrates in an unstirred chemostat; [16] models growth with a limiting substrate that is not broken down after uptake into the cells (such as inorganic ions); [17] proposes growth equations where the specific growth rate does not instantaneously adjust to changes in the concentration of limiting substrate in the chemostat, in place of the simpler Monod growth relation; [18], [19] model delayed growth; the work by Kovarova et. al. [20], [21] expands the Monod growth curves to include elements such as a non-zero minimal limiting substrate concentration below which there is no growth (see section D2 below). Further studies include, among others, [22]–[27].

Within these theoretical studies, there are often "clusters" of papers, where multiple studies research some focused topic. For example, several studies (theoretical or experimental, such as [28]) show that in the "basic" case coexistence of two competing strains in a chemostat with time-invariant operating conditions is practically impossible. There are several other mathematical studies that try to find specific conditions where multiple strains can coexist in steady state: [29] shows a case where stable coexistence could be explained by differential patterns of the secretion and uptake of two alternative metabolites; [30] explores a setup with two complementary substrates in configurations of interconnected chemostats for the coexistence of three competing microbial populations; [31] tries to achive coexistence of three strains by having a periodically varying dilution rate; [32] shows a case where using a non-spacially-uniform environment (namely, two interconnected chemostats) can allow coexistance. There are many other papers on the subject, such as [33]–[35]. There are similarly other "clusters" of papers on focus subjects.

There are other relevant works that were not done directly in connection to chemostats. Basic growth models include the famous Monod study [1], and extensions to his model (for example [36] and the aforementioned papers above). Long term cell cultivation is studied, for example, in the Lenski experiments [37]–[41].

In contrast, in our research we aim to provide a study of the entire chemostat selection process, end-to-end. We model the dynamics and analyze the influence of the various parameters in each stage addressing what we think are the primary issues, while on the other hand focusing in depth on certain interesting points. This includes an explanation of the main parameters and their relationships (section A), analysis of the processes occurring during steady state growth (section B), mutation and evolution (section C), competition, selection and takeover (section D), and the long-term repetitive process required to reach the final selection goal (section E).

The reason for the choice to study these issues, as mentioned above, is so that this study can hopefully be used as an aid for the setup of chemostat experiments, and as a tool for an informed choice of experimental parameters, and ultimately to advance a deeper understanding of chemostat dynamics.

# Methods

The methods we used to answer the various questions range between analytic derivations, computer simulations (implemented using Matlab), and the interplay between the two.

The analytic component includes a modeling of the various stages of the growth and takeover process. This is analyzed using general parametric equations and derivations thereof.

The computational component includes a set of simulations:

- **A numerical visualization tool**, which plots the various curves (Monod, growth, limiting substrate concentration, takeover time, etc.) for the chemostat and cells strains, depending on a list of parameters.
- **A single strain time-stepped growth simulator**, which calculates the differential equations of cell growth and limiting substrate depletion. The growth is determined dynamically by the Monod curve of the strain, for the current limiting substrate concentration. The calculations are done for each dilution pulse, pulse by pulse, where the output of one step is translated into the input for the next step.
- **A two-strain time-stepped competition simulator**, similar to the previous tool, but with the ability to input parameters of two strains, and have them compete for the same diminishing limiting substrate. The simulation begins with a dominant reference strain filling the chemostat, and a single mutant cell. It continues until the mutant strain takes over the chemostat population.
- **Evolution and successive takeovers,** which allows for successive mutant strains to arise at certain conditions and take over the chemostat. The simulation is described in depth in section E.

As is apparent further on, the analytic component is predominant in the justifications of the answers in this manuscript . The results of the computer simulations do not constitute a "proof" for the answers, but rather are used to create plots and graphs to help illustrate the main findings. The simulations also helped "point us in the right direction", in deriving and understanding the analytic proofs. Thus, the interplay between the analytic component and the computational component consisted of a positive feedback loop: The computational results help gain insight and further the analytic structure; the expanded analytic structure was then implemented into the computational tools, and so forth.

## General case & Default characteristic case

We shall assume, quite broadly, only that the reference and mutant strains are self-replicating organisms that adhere to a generalized Michaelis-Menten, or Monod Curve, relation between the growth rate and one limiting substrate (Figure 2 above). The Monod curve for the reference strain is:

$$\mu_0(S) = \frac{\dot{X}_0}{X_0} = \mu_0^{max} \cdot \left(\frac{S}{S + K_0}\right)$$

The "generalized" Monod curve we use for the mutant strain is:

$$\mu_1(S) = \frac{\dot{X}_1}{X_1} = (\mu_1^{max} - \mu_1^{min}) \cdot \left(\frac{S}{S + K_1}\right) + \mu_1^{min}$$

Where $\mu$ is the growth rate, $X$ is the cell concentration, $\mu^{max}$ is the maximal growth rate, $K$ is the Monod parameter, and $S$ is the limiting substrate concentration. The index 0/1 denotes the reference strain and mutant strain, respectively. In the mutant "generalized" form, $\mu^{min}$ is the growth rate when $S = 0$, representing a strain which can grow in media with no limiting substrate (presumably by consuming a different substrate present in the media; we assume that the reference strain cannot do this).

The strains of course need other substrates as nutrients, but we are assuming a domain in which all other substrates are abundant enough so as to not limit the growth rate.

For simplicity we assume the chemostat is "fully mixed", and that the growth rates are instantaneously regulated by the limiting substrate concentration, whereas in reality processes such as diffusion are not instantaneous. As we shall see, the steady state chemical variability is small and dynamics slow, so these assumptions should be sufficient.

Moreover, although we assume a specific type of growth curves (relations between the growth rate and one limiting substrate, in our case – the Monod curve) the results are expected hold when assuming other similar growth curves (a specific example, the Korvarova extension to the Monod curve[20], [21], is examined in the text).

Additionally, we shall calculate and graph the results for the parameter values of the specific default characteristic case of a selection process, as described in [3] of a strain of *E.coli*, which can grow independently of xylose, as was evolved through selection from a reference strain that is xylose dependent.

# III. Results – questions with short answers

## A) Parameter choice and setup

In this chapter we examine the primary elements relevant for the initial setup of a chemostat experiment. The basic parameters of the chemostat and the relationships between them are explored: Chemostat tank volume $V_C$, Dilution pulse volume $V_d$, Dilution interval $\tau_d$ and Dilution parameter $D$ (section A1); Limiting substrate influx concentration in the chemostat $S(t)$ and in the influx drop $S_d$ (section A2); Cell concentration $X(t)$ and OD (section A3); number of cells $N_{cells}$ (section A4); and average steady state growth rate $\mu_0^{st}$ and Dilution parameter $D$ (section A5 and later in chapter "D – Single takeover"). Also, the initialization process from inoculation to steady state growth is analyzed. The chapter gives a first glimpse into chemostat behavior and dynamics, along with laying a foundation for the following chapters and steps.

### A1) How is the "continuous" dilution of an ideal chemostat achieved in practice with dilution pulses?

No chemostat is truly continuous. Every $\tau_d$ seconds a drop of volume $V_d$ is added to the chemostat, and a corresponding volume is removed through overflow.

There is a balance between $\tau_d, V_d$ the chemostat volume $V_C$ and the dilution rate $D$:

$$D = \frac{V_d}{V_C} \cdot \frac{1}{\tau_d} = (1 - \phi) \cdot \frac{1}{\tau_d}$$

Where $(1 - \phi) = \frac{V_d}{V_C}$ is the fraction of media removed in each dilution pulse. For a given $D$ and $V_C$ the dilution can be generated by small frequent drops (small $V_d$ and fast $\tau_d$, Fig. A1 right), or by larger but more infrequent influx pulses (large $V_d$ and slow $\tau_d$, Fig. A1 left). Note, an experiment involving frequent serial dilutions of a batch culture can be viewed as a "pulsed" chemostat experiment.

The implications of pulse size are explored in section B4.

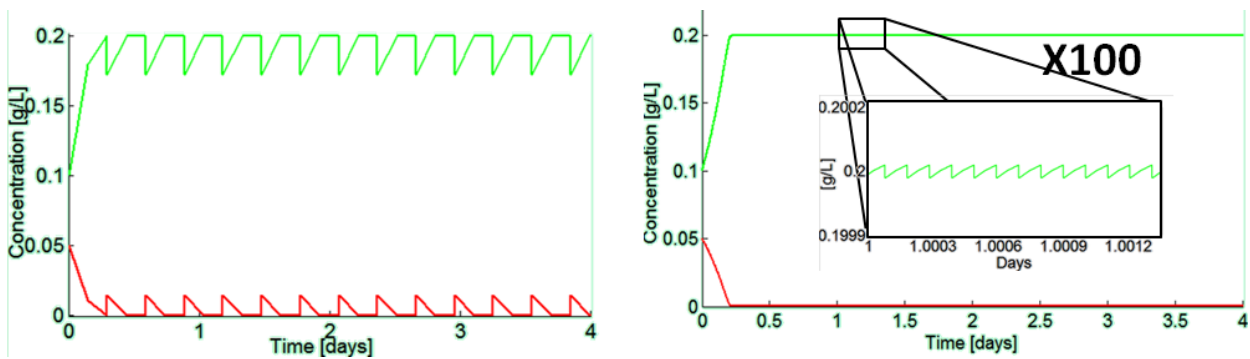

**Fig. A1**: Cell Concentration [X, green line] (OD equivalent of 0.5) and limiting substrate [S, red line] vs. time. [LEFT] A chemostat with very large influx drops, $(1 - \phi) = 15\%$ of the chemostat volume. [RIGHT] A chemostat with our default values $(1 - \phi) = 0.02\%$ (right). The latter is also not truly continuous, because the influx still consists of discreet drops, as seen in the 100-fold zoom box.

Experimentally, $V_d$ is determined by the diameter of the influx tube (which in a characteristic setup is around 5 mm), and is usually around $100\mu L$ (as is our default characteristic value, as shown in fig. A1 on the right). $V_C$ usually does not change during the experiment. $D$ is the main experimental parameter, and the choice of $D$ determines $\tau_d$. $D$ equals the average growth rate (see sections A5 and B5), and can vary between $D=0$ (no growth at steady state) to the maximal growth rate of the cells, above which all cells in the chemostat get washed away.

An advantage of operating the chemostat using drops is that it helps avoid contamination. Bacteria travel upstream quite easily, so to avoid contamination the influx reservoir media should not come in contact with the media in the chemostat vessel, but rather individual drops should fall through the air from an influx tube.

**A2) What is the steady-state concentration of the limiting substrate within the chemostat?**

The steady state limiting substrate concentration in the chemostat is: $S_i^{st} \approx \frac{K_i D}{(\mu^{max}-D)}$ when dominated by cell strain i, as shown in Fig. A2. We note the non-intuitive fact that this value is independent of the input limiting substrate concentration ($S_d$). This means that a higher input concentration does not translate into a higher steady-state limiting substrate concentration, but rather into a higher OD (as is shown in the following section A3). Moreover, since $K_i$ (K Monod of cell strain i) and $\mu^{max}$ (maximal growth rate) are parameters of the strain, the only parameter of the chemostat system that affects the limiting substrate concentration is D.

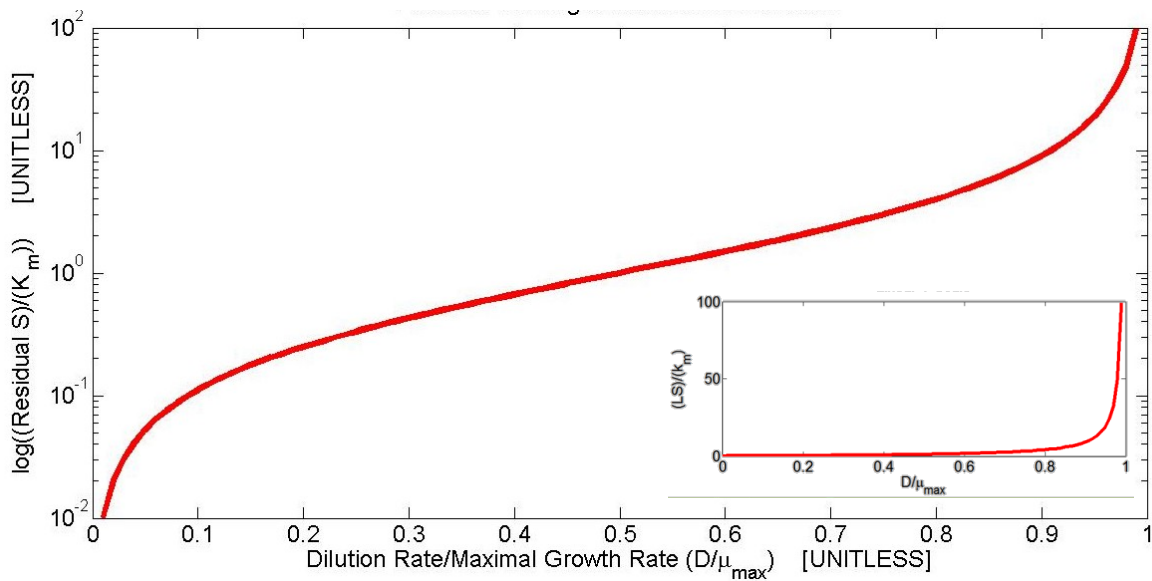

*Fig. A2*: Normalized residual limiting substrate (s) [log scale] vs. normalized dilution rate (D). Inset figure- linear scale.

## A3) What are the steady state cell concentration and OD?

The steady state cell concentration level in the chemostat is: $X_i^{st} = \left(S_d - \frac{K_0 \cdot D}{(\mu^{max} - D)}\right) \cdot Y_0$ as shown in Fig. A3.

Note, the units used for the cell concentration, $X_i^{st}$, are $\left[\frac{g}{L}\right]$ so as to be congruent with the concentrations of substrates. To convert from concentration to OD, a conversion factor $1[OD] = \frac{conversion}{factor} \left[\frac{g}{L} CDW\right]$ must be used. For *E. coli* we use $1 \cdot [OD_{600}] = 0.4 \left[\frac{g\ CDW}{L}\right]$ (see appendix for section A3) so that $OD_0 = \left(S_d - \frac{K_0 \cdot D}{(\mu^{max} - D)}\right) \cdot \frac{Y_0}{0.4}$. We note that the steady state cell concentration is strongly dependent (almost proportional) to the input limiting substrate concentration ($S_d$), unlike the steady state limiting substrate ($S_0^{st}$) (see A2 above). Furthermore, since $S_0$ is dependent on only one system parameter ($D$), and $X_0^{st}$ is dependent on two system parameters ($D, S_d$), $S_0^{st}$ and $X_0^{st}$ (the limiting substrate internal concentration and steady state concentration of the cells) can be independently set.

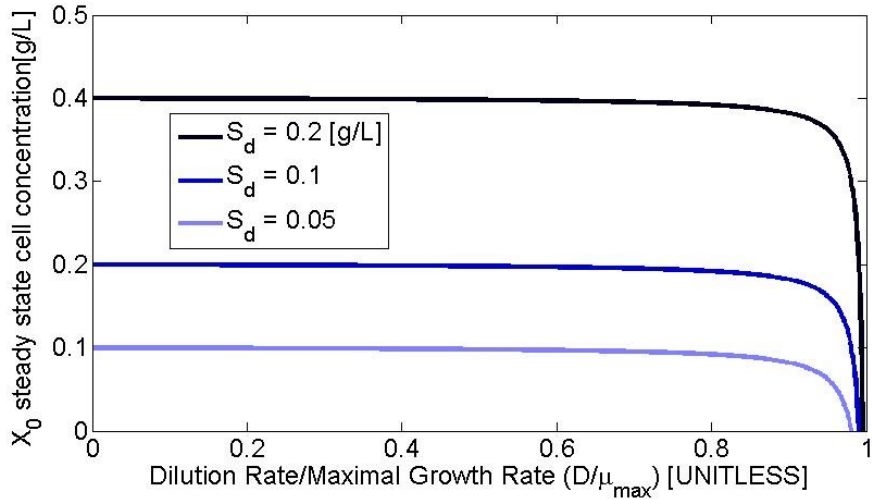

***Fig. A3***: *The steady state cell concentration, as a function of the dilution rate (D), for different input limiting substrate concentrations ($S_d$). The cell concentration is almost constant, except for dilution rates close to the maximal growth rate, where washout is approached.*

## A4) What is the number of cells in the chemostat at steady state?

The number of *E.coli* cells in the chemostat vessel at steady-state, (assuming an OD of 1 is equivalent to about $10^9 \frac{cells}{mL}$ or $10^{12} \frac{cells}{L}$), is:

$$N_{cells} \approx OD_0 \cdot V_C \cdot 10^{12} = \left(S_d - \frac{K_0 \cdot D}{(\mu^{max} - D)}\right) \cdot \frac{Y_0}{0.4} \cdot V_C \cdot 10^{12}$$

Where $Y_0$ is the yield, and on the right-hand side we used the ratio of $1 \cdot [OD_{600}] = 0.4 \left[\frac{g\ CDW}{L}\right]$. In our case the number of cells is (assuming our default characteristic values for the other parameters):

$$N_{cells} \approx 3 \cdot 10^{11}$$

For comparison, the number of budding yeast cells (with similar cell growth parameters, such as $Y_0, K_0, \mu^{max}$) is:

$$N_{cells} \approx OD_0 \cdot V_C \cdot 3 \cdot 10^{11} = \left(S_d - \frac{K_0 \cdot D}{(\mu^{max} - D)}\right) \cdot \frac{Y_0}{0.6} \cdot V_C \cdot 3 \cdot 10^{11}$$

$$N_{cells} \approx 5 \cdot 10^{10}$$

## A5) What is the average cell growth rate and doubling time at steady state? When should a factor of ln(2) be used?

Going back to the definition of the dilution rate D, it equals "the fraction of the chemostat volume replaced per hour by inflow from the reservoir". Therefore, the differential equation for the cell concentration, accounting for growth and dilution, is (as shown by [5]):

$$\frac{dX(t)}{dt} = \mu \cdot X(t) - D \cdot X(t)$$

At steady state, $\frac{dX(t)}{dt} = 0$, so that the average growth rate of all the cells in the chemostat ($\mu^{st}_{average}$) equals the dilution rate ($D$).

In other words, every dilution pulse removes $V_d * X_0^{st}$ cells which are replaced exactly by growth over the next interval (of duration $\tau_d$). This is the definition of steady-state: because the system is at steady-state dilution and growth must balance exactly to have no net change in the cell concentration $X_0^{st}$.

The connection between dilution rate and doubling time can be calculated as follows:

Growth rate is defined as: $\dot{X} = \mu \cdot X$. The solution is: $X = X_{t=0} \cdot e^{\mu t}$. Solving for the doubling time ($t^{doubling}$):

$$2 * X_{t=0} = X_{t=0} \cdot e^{D * t^{doubling}}$$

$$t^{doubling} = \frac{\ln(2)}{D}$$

This provides us with a seeming discrepancy between the turnover time 1/D, and the doubling time.

On the one hand, naively, since the cell concentration remains constant over time, the cell population must have doubled over 1/D hours (in this time the entire volume of the chemostat vessel is replaced by dilution, as this is the definition of the dilution rate) - $V_c * X_0^{st}$ cells were removed and $V_c * X_0^{st}$ cells remain in the growth vessel. The growth that occurred in this time frame should thus be equivalent to one doubling of the cell population.

But, on the other hand, calculating from growth rate, we expect the time required to be shorter - $\frac{\ln(2)}{D}$.

In other words, if the population doubled after $\frac{\ln(2)}{D}$, and after 1/D there are still only $2 * V_c * X_0^{st}$ cells – where are the "missing cells" that should have grown in the time difference $\frac{1-\ln(2)}{D}$?

The solution lies with the fact that the waste cells continue to grow after being removed from the chemostat.

For proof see appendix A5.

(Also see section B5 and corresponding appendix B5 for further analysis of the growth rate).

### A6) How does the initial amount of limiting substrate in the chemostat affect the setup time?

Typically, the chemostat is first filled with sterile media and then inoculated with cells. Usually, the initial concentration of limiting substrate ($S_{init}$) is substantially higher than the steady state limiting substrate concentration in the chemostat (or $S_{initial} \gg S_0^{st}$) (see section A2). After inoculation, the cells grow rapidly until the initial limiting substrate ($S_{init}$) is depleted, and only then the chemostat dynamics (i.e. substrate limitation, dilution, etc.) starts to be significant.

The total mass of atoms originating from the limiting substrate is conserved, in one form of other. Those atoms can either still be in their original form as free residual limiting substrate (i.e. residual xylose, $S_0^{st}$), part of the cells (i.e. biomass carbon originating from the xylose), or converted into some byproduct (i.e. CO2). In steady state the total mass of atoms originating from the limiting substrate mass in the chemostat equals the mass in a chemostat volume worth of limiting substrate influx concentration $S_d * V_c$.

Thus, if too much limiting substrate mass is initially present ("free" in a high initial concentration $S_{init} > S_d$, or in the inoculation cells when the inoculum size is substantial) – the OD will rise above steady state levels ("overshoot", Fig. A6 left). If too little limiting substrate is present – the OD will be below steady state levels when the initial limiting substrate is depleted, and the chemostat dynamics take over ("undershoot", Fig. A6 right). In both cases, the time it takes to reach steady state is prolonged, where:

$$OD(t) = (OD_{peak} - OD_{SS}) \cdot e^{-Dt} + OD_{SS}$$

Meaning that the difference in OD is reduced by a factor of $e^{-1}$ every chemostat turnover. The prolonged time is on the order of several chemostat turnovers.

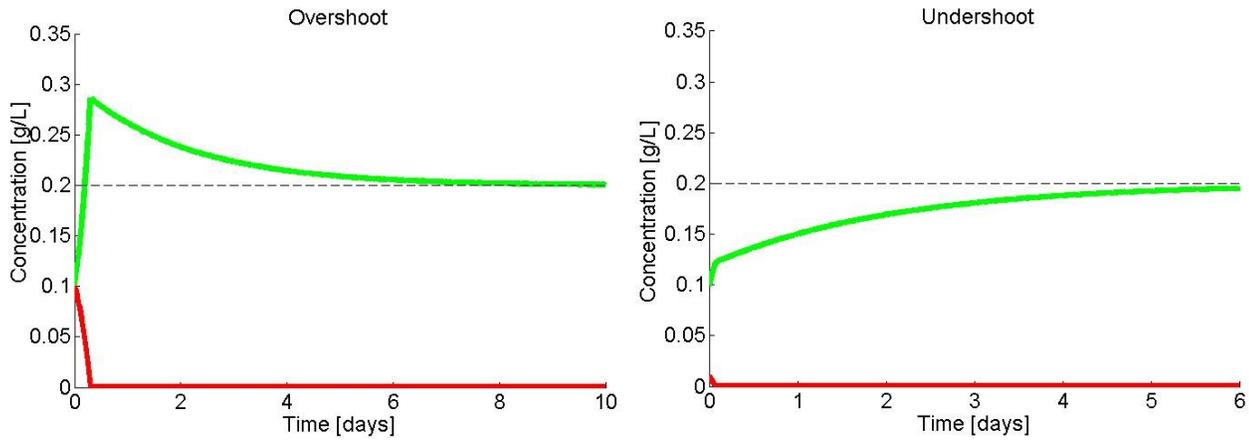

***Fig. A6:*** *Cell Concentration [X, green line] and limiting substrate [S, red line] vs. time. Black line represents steady state OD. Overshoot (left) and Undershoot (right).*

## A7) What is the "Optimal boot sequence" for chemostat setup?

The fastest way to achieve steady state is by first allowing the cells to reach their steady state concentration value as fast as possible, and then allow the cells to remain in this concentration, thus circumventing any overshoot/undershoot (see section A6 above).

The first step can be achieved by growing the chemostat in "batch" (i.e., with no dilution slowing down the growth). Once reaching steady state OD (which is determined mostly by the amount of limiting substrate in the feed and more weakly by the desired dilution rate *D,* see section A3) dilution should begin. The circumvention of overshoot/undershoot can be achieved by choosing an initial limiting substrate concentration in the chemostat to exactly bring the chemostat optical density value to the steady state value, or:

$$S_{initial} = S_d - \left(\frac{Cell\ Concentration_{initial}}{yield_{reference}}\right)$$

where the $Cell\ Concentration_{initial}$ is measured in relation to the entire chemostat volume, $V_C$. $S_{initial}$ is usually very close to $S_d$. Specifically, when the inoculum is very small the chemostat can be initially filled from the influx reservoir at concentration $S_d$.

This careful choice of concentration achieves perfect matching. The procedure can alleviate the overshoot/undershoot that results from having dilution from the start and that results in a relatively long delay in reaching steady state on the order of days to weeks depending on dilution rate. One should be careful not to wait after achieving the OD as this will lead to cell starvation.

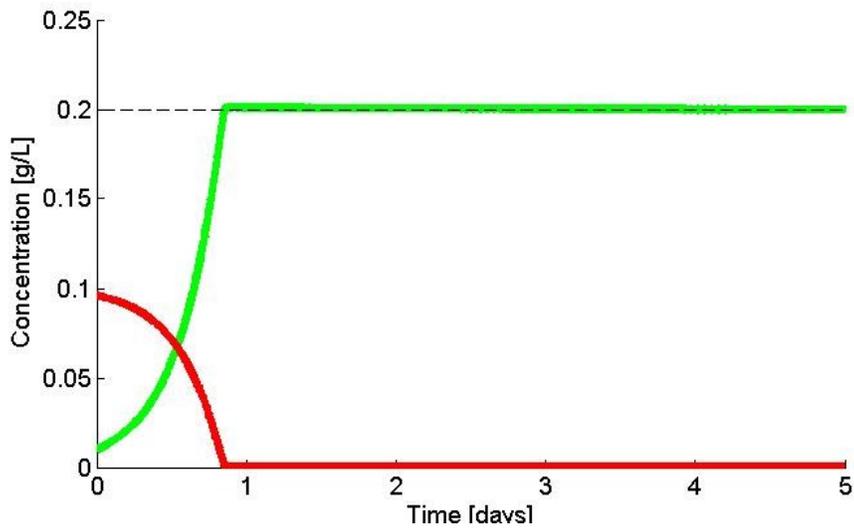

***Fig. A7***: *Cell Concentration [X, green line] and limiting substrate [S, red line] vs. time. Black dotted line represents steady state OD. An example of careful choice of initial limiting substrate and cell concentrations so that they reach desired steady state values simultaneously. Thus, overshoot and undershoot are avoided.*

# B) Basic "steady state" growth

## B1) Is the system really in steady state? What is steady, what isn't?

The meaning of "steady state" for our purposes is from an experimental point of view, where the macroscopic dynamics of the chemostat remain constant. No chemostat is perfectly continuous since the temporal dynamics following each influx drop are not steady. For example, the OD drops slightly because a drop of media has been added, and then the cells grow, etc. Therefore, technically, the macro-dynamics are also not constant, but "constantly repetitive" over a long period of time, i.e. the dynamics following each influx drop repeat, as shown in Fig B1. Note that for our characteristic parameters the value of $\Delta X$ is less than 1/1000 of $X$ and $\tau_d$ is ~10-100 seconds. In these circumstances, the sawtooth pattern in figure B1 will usually not be detectable. If the variability in concentrations is smaller than the resolution of detection then the concentrations will appear constant. We chose here to zoom in for didactic reasons.

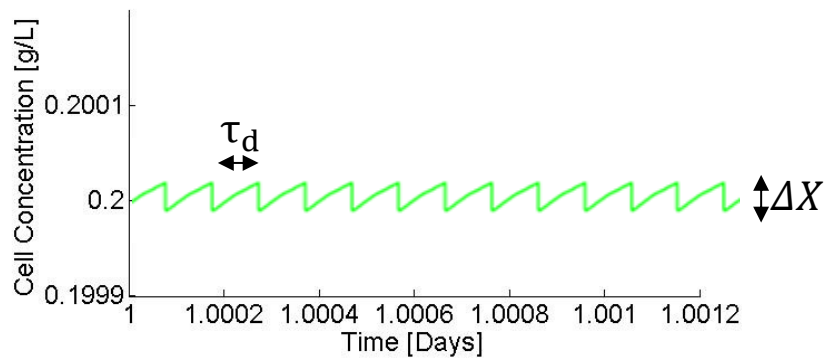

***Fig. B1***: *Zoom in on cell concentration [green line] dynamics in chemostat at steady state. On a short time scale, $\tau_d$, the chemostat is not "steady" as the concentration is not constant but rather goes down and up. The dynamics are constantly repetitive in that the dynamics following each drop stays the same. Note that for characteristic parameters the value of $\Delta X$ is less than 1/1000 of X and that $\tau_d$ is on the order of a minute and thus the changes that can be observed in the figure will usually not be detectable. If the variability in cell concentration, $\Delta X$, is smaller than the detection level then the concentration will appear constant. We chose here to zoom in for didactic reasons.*

**What is relatively steady:** The OD level, limiting substrate concentration, most other non-limiting substrates concentrations are in steady state, until a specific kind of strain takeover occurs (see following section B2).

**What is not steady:** Strain-wise, the chemostat is very dynamic. There could be many competing strains, at low frequencies, and a few at higher frequencies (thus detectable by sequencing), all constantly changing. There can be many successive sweeps (takeovers) that are non-detectable at the OD or substrate levels (see section E).

## B2) What strain takeovers are detectable by monitoring OD or residual substrate levels in the chemostat?

Most mutant takeovers in the chemostat are not detectable at the macro level (i.e. without molecular tests like sequencing). Rather a new strain will grow to dominate the population without markedly changing the macro properties of the chemostat enough to be noticed over the detection noise. To clarify, the undetectable takeovers are not only neutral drift or fitness improvements that are too small to be of significance, but rather even substantially improved strains. We now describe types of takeovers that do change the macro properties and are detectable. For further elaboration see [42].

**OD change:** if a strain that creates more biomass per input limiting substrate (i.e. generating higher biomass yield, or using an alternative substrates instead of the limiting substrate) becomes dominant, the OD should rise. For example, in the experiment of Antonovsky et al.[3], when evolution led to a strain that could fully utilize $CO_2$ to make sugar, the originally limiting substrate xylose was not required anymore and the high abundance of pyruvate led to a sharp increase in OD. (See Fig. B2, LEFT)

**Diminishing limiting substrate:** often the limiting substrate is at undetectably low values (especially at slow dilution rates). If the limiting substrate is detectable, and a strain that is better at uptaking the limiting substrate (better effective $K$ Monod) becomes dominant, the limiting substrate should drop to a new lower steady state value (and the OD should rise somewhat but usually not by a detectable extant). (See Fig. B2, CENTER)

**Diminishing non-limiting substrates:** substrates that are not growth limiting are usually found at detectable levels in the chemostat vessel. A decrease in one of these substrates can indicate a strain takeover of one of two kinds:

The first is a strain that consumes the non-limiting substrate without converting it into biomass (see Fig. B2, RIGHT). In the experiment of Antonovsky et al.[3], a mutant strain that processed pyruvate into acetate became dominant, and lowered the pyruvate levels substantially.

A second type of takeover that is detectable as a diminishing non-limiting substrate concentration is by a strain which replaces the use of the limiting substrate (e.g. xylose) with a previously non-limiting substrate (e.g. pyruvate) that becomes limiting (see Fig. B2, RIGHT). This is not necessarily detectable as a change in the OD (as above).

Note that since the biomass production is limited by the limiting substrate (i.e. xylose), a mutant strain with a more efficient uptake of non-limiting substrate (i.e. a higher affinity to pyruvate) will not be detectable as a change in OD.

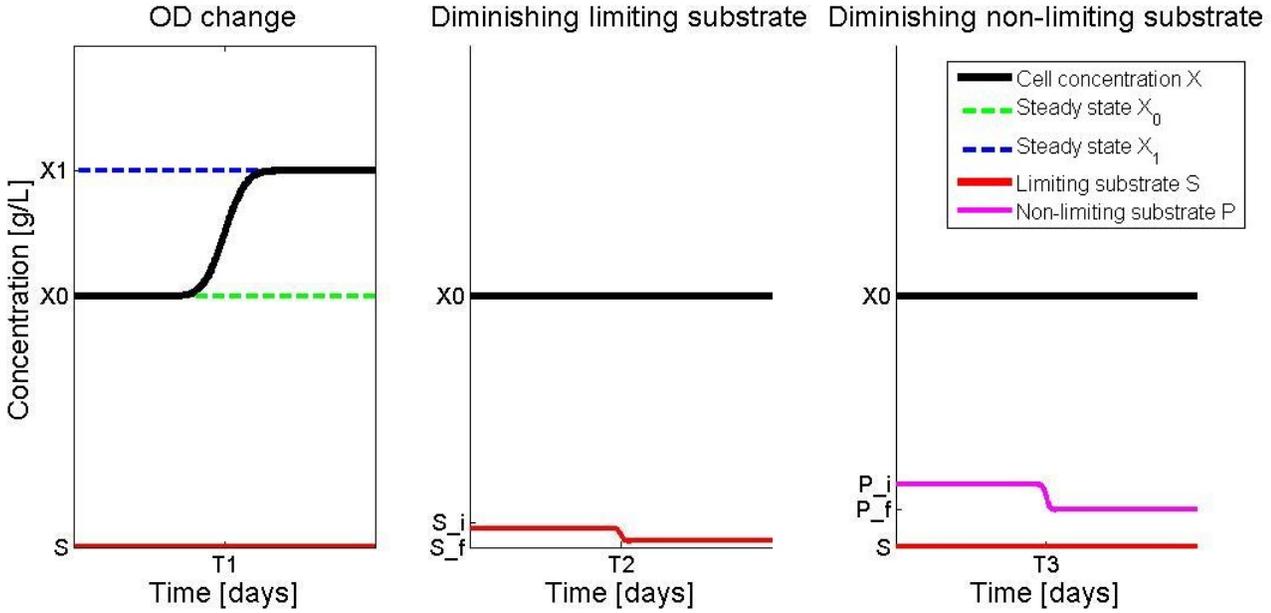

*Fig. B2: Three types of detectable strain takeovers are depicted. **(LEFT) Cell concentration change**: [black line rising from X0 to X1] takeover of a strain that creates more biomass per input limiting substrate (higher yield, or can use alternative substrates instead of limiting substrate). **(CENTER) diminishing limiting substrates**: [red line, dropping from Si to Sf] can reveal the takeover of a strain better at uptaking the limiting substrate S. This is not necessarily detectable in an OD change if the change is smaller than the OD measurement resolution. **(RIGHT) diminishing non-limiting substrates**: [red dotted line, dropping from Pi to Pf] can occur either due to (1) takeover of a strain that reduces the substrate without uptaking it into biomass; or (2) a strain that uses P instead of S.*

### B3) What is the temporal variation in limiting substrate?

As we shall see (in section B4), a relatively stable limiting substrate concentration is usually desired. Since the limiting substrate levels in the chemostat are often very low (see section A2), a drop of influx can lead to a large change in the limiting substrate concentration, i.e. large temporal variability. In practice, the variability in limiting substrate, or $\Delta S = S_{after\,drop} - S_{before\,drop}$ is (using the value for $S_0^{st}$ derived in section A2):

$$\Delta S = \frac{V_d}{V_C - V_d} \cdot (S_d - S_0^{st}) = \frac{V_d}{V_C - V_d} \cdot \left(S_d - \frac{KD}{\mu^{max} - D}\right)$$

Where $\left(\frac{V_d}{V_C - V_d}\right) < 1$. For our default characteristic case the absolute variability is $\Delta S = 2 \cdot 10^{-5} \left[\frac{g}{L}\right]$. The relative variability is:

$$\frac{\Delta S}{S_0^{st}} = \frac{V_d}{V_C - V_d} \cdot \left(\frac{\mu^{max} - D}{KD} S_d - 1\right)$$

For our default characteristic values the limiting substrate concentration prevailing within the chemostat is $S_0^{st} = 7 \cdot 10^{-4} \left[\frac{g}{L}\right]$, so that the relative variability is only about $\frac{\Delta S}{S_0^{st}} \approx 3\%$.

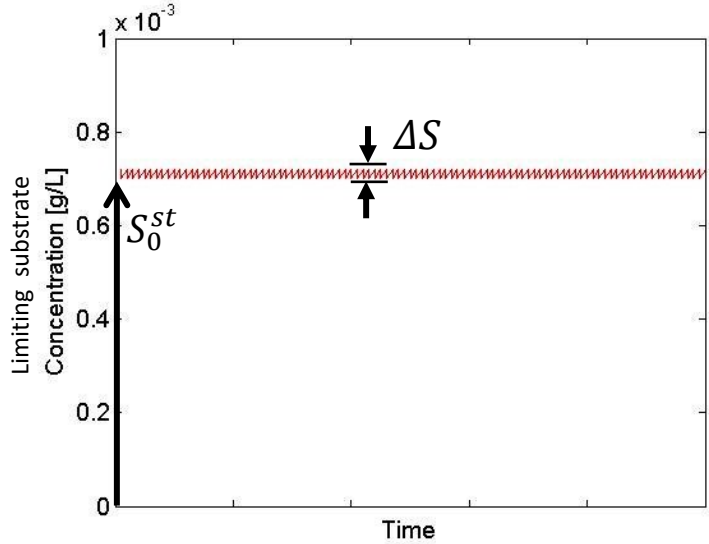

***Fig. B3:*** *Variability in limiting substrate concentration due to dilution, for characteristic default values. The variability is small in comparison to the steady state concentration, or $\Delta S = 3\% * S_0^{st}$.*

From the equation above for $\frac{\Delta S}{S_0^{st}}$ it is apparent that:

- A larger amount of limiting substrate in the influx pulse (either via a larger $V_d$ balanced by a smaller $\tau_d$ to keep D constant, or via a larger $S_d$) results in a larger absolute variability $\Delta S$, and a larger relative variability $\frac{\Delta S}{S_0^{st}}$, (as the influx limiting substrate concentration does not change the steady state concentration $S_0^{st}$).

- When the dilution rate D is slowed, approaching D=0, the limiting substrate influx is practically used up between drops. The steady state concentration is will also be close to 0, so that $S_0^{st} \underset{D\to 0}{\to} 0$. Thus the absolute variability is the entire influx drop, $\Delta S \underset{D\to 0}{\to} \left(\frac{V_d}{V_C - V_d}\right) \cdot S_d$, and the relative variability is unbound $\frac{\Delta S}{S_0^{st}} \underset{D\to 0}{\to} \left(\frac{\Delta S}{0} = \infty\right)$.

- Since the steady state limiting substrate concentration is directly proportional to the substrate affinity K, a smaller K (more efficient substrate uptake) produces a lower steady-state concentration $S_0^{st}$, resulting in a larger variability in the concentration. For instance, for our default characteristic values $\frac{\Delta S}{S_0^{st}} \approx 3\%$, but for a smaller K value of $K_{new} = 0.1 K_{default}$ we get $\frac{\Delta S}{S_0^{st}} \approx 30\%$. Vice versa, a higher K value results in a smaller variability.

- When D approaches $\mu^{max}$ the variability goes to zero $\frac{\Delta S}{S_0^{st}} \underset{D\to \mu max}{\to} 0$. This is because a fast dilution rate (D) leads to relatively high limiting substrate levels (approaching the influx concentration $S_D$) and fast uninhibited growth (section A2). In this case, substrate depletion due to uptake is negligible.

**B4) How does a pulsed chemostat differ from a quasi-continuous one?**

For technical reasons, a chemostat with a perfectly continuous influx is not possible, and most chemostats deliver influx as discrete drops of fresh media. A "pulsed chemostat" delivers relatively larger influx drops at longer time intervals (fig. B4 left) than a "quasi-continuous chemostat" (fig. B4 right). There is no clear cutoff criteria for determination that a chemostat is one regime or the other.

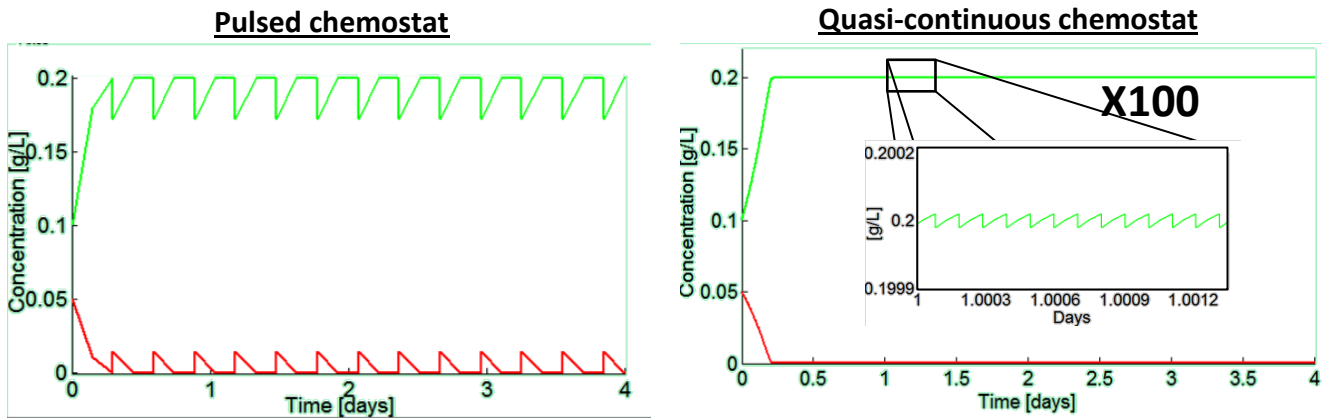

*Fig. B4:* Cell Concentration [X, green line] (OD equivalent of 0.5) and limiting substrate [S, red line] vs. time. Pulsed chemostat with a pulse size of $(1 - \phi) = 15\%$ of the chemostat volume (left) and Quasi-continuous chemostat with our default values $(1 - \phi) = 0.02\%$ (right). The latter is only "Quasi" continuous because the influx still consists of discreet drops, as seen in the 100-fold zoom box.

Whether the chemostat operates at one or the other regimes is important because in the "pulsed" chemostat there is a substantially lower selective advantage for a mutant strain with increased fitness in limiting conditions. Qualitatively, this is because just after each influx drop the cells are subject to a relatively high concentration of limiting substrate and grow quickly until the limiting substrate is depleted. Hence the "pulsed" dynamics gives a lower advantage to the mutant cells. In the quasi-continuous case the concentrations stay relatively stable, giving a steady selective advantage.

Using numerical simulations, it can be shown that sufficiently large influx drops are enough to substantially delay mutant takeover in a non-linear fashion. For our default characteristic values, when all other parameters are held constant, using drops of around $(1 - \phi) \approx 2\%$ (instead of the nominal 0.02%) almost does not change the takeover time, while using drops of 4% prolongs takeover time by a factor of 1.7.

**B5) What is the calculated average cell growth rate of the reference and mutant strains at steady state?**

In section A5 above we determined the average growth rate of all the cells in the chemostat considering the time interval during which the chemostat volume is replaced. Here, and in appendix B5, we calculate the average growth rate of each strain based on the limiting substrate concentration, (as dictated by the dominant strain see section A2) and the Monod growth relations.

The average growth rate for of the reference strain is $\mu_0 = D$ when the chemostat is dominated by this strain. The average growth rate of the mutant strains, when the chemostat is dominated by the reference strain, is:

$$\mu_1(D) = \frac{\mu_1^{max} D + \mu_1^{min} \frac{K_1}{K_0}(\mu_0^{max} - D)}{D + \frac{K_1}{K_0}(\mu_0^{max} - D)}$$

**B6) What happens when the limiting substrate influx is abruptly lowered?**

In the long run, as discussed above in sections A2 and A3, the steady state cell concentration (and OD) will change, but the limiting substrate will not. The transient stages, until return to steady state, are shown in Fig. B6, as follows:

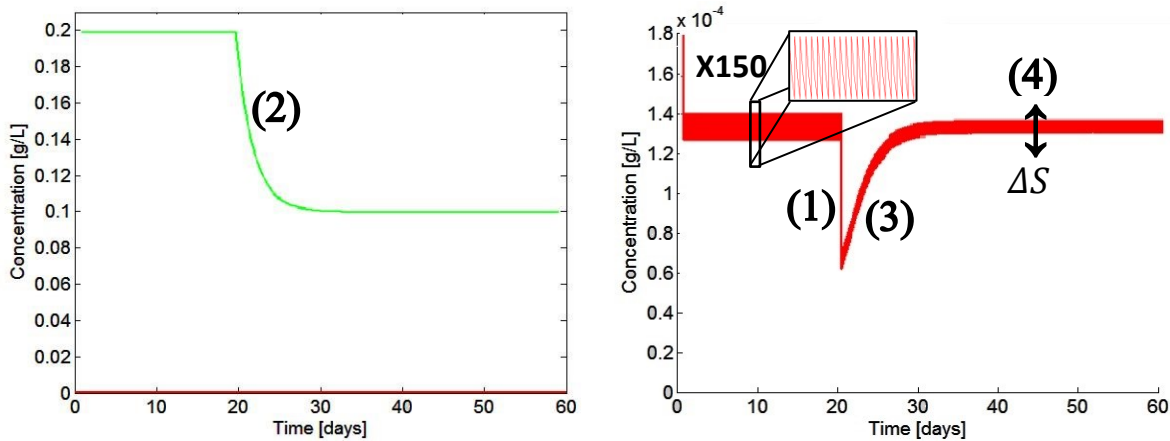

***Fig. B6:*** *(**LEFT**) changes in cell concentration (X, and therefore OD) (green) and limiting substrate concentration (S) (red) after limiting substrate influx concentration (Sd) lowered by a factor of 2 at Day=20. (**RIGHT**) zoom-in of the y-axis on the red line from the figure on the left. (**INLAID**) a further zoom-in on the X-axis to show individual dilution pulses (drops).*

(1) The limiting substrate in the chemostat drops by a factor close to that which the input limiting substrate was lowered (for example, in Fig. B6 the input limiting substrate was lowered by a factor of 2 and then the limiting substrate in the chemostat drops by a factor of 2). This happens on a very fast time scale (minutes).

(2) This leads to a decay in the OD, to the new steady state on a time scale of 1/D; where the difference of OD between steady-states decays proportionally to $e^{-Dt}$.

(3) This leads to a corresponding rise in the limiting substrate concentration from the transient lower value back up to the new steady state concentration, which occurs over the same time scale that the OD reaches its new steady state (steps 2-3 happen simultaneously).

(4) Since the OD and $S_d$ have changed – the variability in limiting substrate ($\Delta S = S_{after\ drop} - S_{before\ drop}$, which exists because the chemostat is not perfectly continuous) changes slightly (as discussed in section B3).

Symmetrically, if the limiting substrate influx is increased, the limiting substrate in the chemostat will first increase, leading to an exponential rise in the OD and an exponential decay of the limiting substrate back to steady state levels.

**B7) What happens when the limiting substrate is constantly and continuously lowered?**

Here the lowering is continuous, as opposed to section B6, where the influx concentration was lowered abruptly. In B6 the change was abrupt, after which the OD and limiting substrate concentration stabilized to a new steady state. In our case, the constant change does not leave time for stabilization, and the limiting substrate in the chemostat continuously goes down. This way the steady state rule we saw above, that D is the only controllable-parameter that dictates the concentration of the limiting substrate in the chemostat, is broken. This phenomenon is partially due to the fact that the adjustment back to steady state is not governed by a direct feedback loop, but rather an indirect feedback loop; the limiting substrate changes the OD that in turn changes the limiting substrate. This indirect loop is relatively slow.

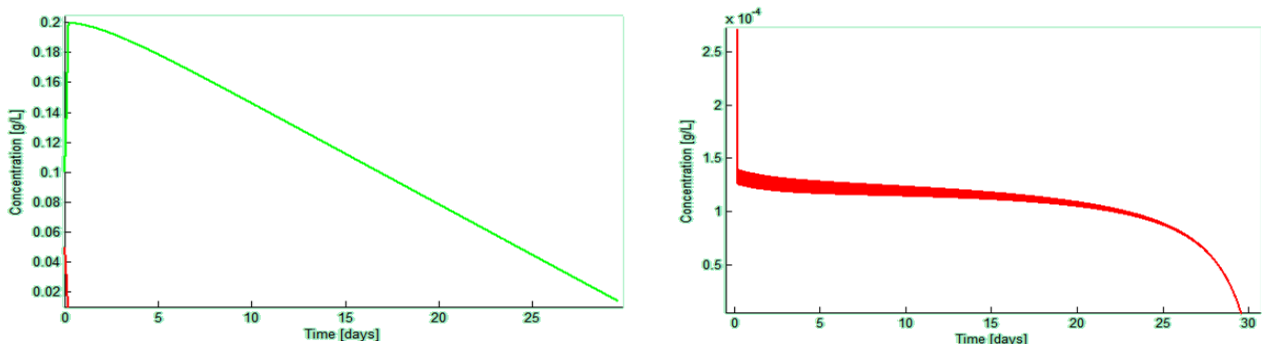

*Fig. B7: changes is Cell Concentration [X, green line] and zoom in on limiting substrate [S, red line] while limiting substrate influx gradually lowered constantly. Right side is a zoom in (y axis only) on the limiting substrate shown at near-zero levels on left side.*

**B8) How do chemostats differ from other similar devices, such as turbidostats?**

The various growth apparatus, such as the turbidostat, auxostat, morbidostat, etc. are similar to the chemostat in that they all allow long-term continuous growth in controlled conditions. At steady state equilibrium they operate mostly the same. However, the other growth apparatus are more complex experimentally than the chemostat, since they require some form of sensory feedback. The added complexity can, on the one hand, enable specific capabilities that cannot be achieved with the simpler chemostat, but on the other hand can generate complications that do not arise in the more robust chemostat.

We shall focus on the turbidostat to exemplify the above points. A chemostat has a fixed dilution rate, while a turbidostat dynamically regulates the dilution rate based on online readings of OD (more formally known as turbidity and hence the name turbidostat). This, of course, requires an added setup for the sensory feedback loop (such as a spectrophotometer).

For both apparatuses, at steady state, the OD stays constant, so that the dilution rate stays constant - the influx and efflux exactly cancel each other out, and the time between drops stays constant. Thus the chemostat and turbidostat at any given time operate the same.

Specific capabilities are enabled in the turbidostat. A dilution rate close to the maximal growth rate of the cells is precarious and delicate in a chemostat, since if the rate is too fast the cells will wash out. In a turbidostat, if the cells begin to wash out the dilution will halt, allowing the population to recover. Yet the sensory feedback generates a risk of biofilms sitting on the OD reader and thus obstructing the control, and of other turbidity factors that can complicate the experiments.

The difference also influences what happens when the cells mutate: the cells can adapt to grow more quickly on the concentration of the limiting factor. In a chemostat the **growth rate** must stay constant and equal to the dilution rate in steady state after the mutant had taken over; for that to happen, the concentration of the limiting substrate will further decrease to "cancel" the faster growth, and at the same time the OD will rise (usually only slightly, see sections B1-B2). In contrast, in a turbidostat the **turbidity** must stay constant; for that to happen, the dilution rate becomes quicker and the limiting substrate concentration may not be affected. In both cases the mutant strain still has a selective advantage and will overtake the ancestor, and both apparatus can be used in selection experiments. When cells mutate and the conversion yield changes – the growth rate does not change, so the chemostat dynamics continues as before. In contrast, in a turbidostat a rise in OD (assuming the yield improved) will trigger faster dilution to decrease back the OD to the set point, thus leading to faster growth.

These are delicate differences, and each type of bioreactor will behave slightly differently in specific situations. For most applications, these differences are not essential, and the simpler chemostat can be used. In certain specialized cases where a specific capability is desired - the other bioreactors can give an advantage (such as choosing a turbidostat when wanting to select for improved maximal growth rate).

# C) Mutation

In this chapter we estimate the time and the number of chemostat doublings needed, on average, for a desired mutated strain to appear. It will be shown that in some cases the desired mutations should appear quickly, and in other cases they are highly unlikely as follows:

A desired mutation of one specific SNP (single-nucleotide polymorphism) should appear after only a few chemostat doublings, for a sufficiently large chemostat ($10^{11}$ cells or above). Also, a strain with two desired SNPs, where only their combination gives a fitness advantage whereas each one separately is neutral, is likely to appear only if the target size (the number of SNP locations that give rise to an advantageous mutation) for each SNP is very large. In other cases (e.g. two SNPs with a small target size, more SNPs or smaller chemostats) the appearance of the desired strain is highly unlikely, and is expected to require so much time to make the attempt ineffective unless each mutation is beneficial on its own.

Additionally, it will be shown that for a large chemostat we expect the time until an advantageous mutation occurs ($T_{mutant}$) to be substantially shorter than the time for an advantageous strain to take over the chemostat population ($T_{takeover}$, as will be addressed in section D), and not necessarily so in a small chemostat.

Note, we shall show calculations for SNP mutations, but similar insights should hold for other types of mutations.

To show these points, the questions will be broken up into a few steps. Note that this chapter discusses the mutation process only. The accumulation of mutations and genetic variation through successive advantageous mutations that take over and become fixed in the population is discussed in the following chapters.

## C1) How many cell doublings, on average, are needed for a specific desired SNP to appear?

Defining the size of the genome of the species to be $G = genome\ size\ [bp]$, and defining the mutation rate per base pair per replication to be $R_{mut} = mutation\ rate\ \left[\frac{SNP}{bp*doubling}\right]$, we get that the number of expected SNPs per cell doubling is:

$$P_{SNP} = G \cdot R_{mut} \left[\frac{SNP}{cell * doubling}\right]$$

For our *E. coli* test case we use $G = 5 \cdot 10^6\ [bp]$ and $R_{mut} = 10^{-10} \left[\frac{SNP}{bp*doubling}\right]$.

The number of individual cell doublings (genome replication events) we need to get the desired mutant strain is approximately:

$$doublings_{cell} \approx \frac{G}{P_{SNP}} = \frac{1}{R_{mut}} = 10^{10} \ [doublings]$$

**C2) How many chemostat doublings, on average, are needed for a specific desired SNP to appear? How is does this depend on the size of the chemostat population?**

The transition from cell doublings to chemostat doublings is conceptually very simple but we perform it in detail to show the major impact of the chemostat size and density. The number of chemostat doublings required depends on the number of single cell doublings required and the number of cells in the chemostat. The number of single cell doublings is analyzed above (section C1).

The number of cells in a standard chemostat is usually between $10^7$ cells for a "small and dilute" chemostat ($V_C = 0.01L, OD_{600} = 0.001 \ E.coli, \ OD_{600} = 0.004 \ Budding \ Yeast$) and $10^{12}$ cells for a "large" chemostat ($V_C = 1L, OD_{600} = 1 \ E.coli, \ OD_{600} = 4 \ Budding \ Yeast$) (see section A4 above). For our default characteristic values $N_{cells} = 10^{12} \cdot V_C \cdot OD_0 = 3 \cdot 10^{11}$.

The number of chemostat doublings needed to achieve $doublings_{cell}$, and therefore the desired SNP, is:

$$doublings_{chemostat} = \frac{doublings_{cell}}{N_{cells}} = \frac{1}{R_{mut} \cdot N_{cells}}$$

and for *E. coli*:

$$doublings_{chemostat} = \frac{1}{R_{mut}} \cdot \frac{1}{OD_0 \cdot V_C \cdot 10^{12}}$$

Using our default characteristic numbers:

$$doublings_{chemostat} = \frac{10^{10}}{3 \cdot 10^{11}} = 0.03$$

Therefore we can expect a mutant strain with one desired SNP to be present in the chemostat after less than one doubling in our default characteristic case, and certainly in the "large" $10^{12}$ cell chemostat.

For a "small and dilute" chemostat of $10^7$ cells the number of chemostat doublings needed is:

$$doublings_{chemostat} = \frac{10^{10}}{10^7} = 1000 \ doublings$$

For a doubling time of approx. 10 hours, we get an impractical amount of time until the one-SNP mutation appears on average, or:

$$Total \ Time \approx 10,000 \ hours \approx 1 \ year$$

**C3) How many cell doublings and chemostat doublings, on average, are needed for a strain with two specific SNPs to appear?**

Single cell doublings

As a reminder, we assume that the two mutations only confer an advantage when present together in the same cell and that they are individually fitness neutral. It can be assumed that the first and second SNP in each strain occur independently, so the number of SNPs required is squared, therefore:

$$doublings_{cell} \approx \frac{1}{2} \cdot \frac{1}{R_{mut}^2}$$

For our default characteristic values

$$doublings_{cell} \approx 5 \cdot 10^{19} \, [doublings]$$

Chemostat doublings

As before, the transition from cell doublings to chemostat doublings depends on the number of single cell doublings required and the number of cells in the chemostat:

$$doublings_{chemostat} = \frac{1}{2} \cdot \frac{1}{R_{mut}^2 \cdot N_{cells}}$$

Using our default characteristic numbers, $doublings_{chemostat} \approx 10^{10}$, meaning it should take over a million years for two specific SNPs to appear in the chemostat.

To understand the vast difference between one and two SNPs, note that the first SNP has a background strain of $N_{cells}$ from which to evolve, but the second SNP in the combination only has the cells that accumulated the first mutation in the first step. This means that the cell doublings required for the second SNP do not occur via many separate cells in parallel as for the first SNP, but via a small number of cells doubling repeatedly in succession.

In conclusion, a mutant strain that requires two (or more) specific SNPs together to gain a selective advantage (where each SNP is non-beneficial on its own) is very unlikely in a chemostat setting. This finding may seem counterintuitive, since in practice mutants appear with more than one SNP. There are several ways a mutant strain with two (or more) SNPs can take over the chemostat population: when only one SNP gives the advantage (the "driver") and the other SNP is random ("passenger"), when each SNP already gives a selective advantage on its own and successive takeover sweeps take place (as will be shown in chapter E), and when the target size for each SNP is sufficiently large (as will be analyzed in the following questions).

## C4) How many cell doublings and chemostat doublings are needed when the target size (θ) for a desired advantageous mutation is larger than one?

Defining target size $\theta$ to be the number of SNP locations that give rise to a certain desired advantageous mutation, the number of cell and chemostat doublings needed to achieve an advantageous mutant strain is:

$$doublings_{cell} = \frac{1}{\theta \cdot R_{mut}} \ [doublings]$$

$$doublings_{chemostat} = \frac{1}{\theta \cdot R_{mut} \cdot N_{cells}} \ [doublings]$$

For our default characteristic values and for a target size of $\theta = 1000 \ [bp]$ (target sizes may vary, we chose a value on the order of the length of a gene, similar to that used in [2] whose experiment resembles our setup, and similar to [43],[44]):

$$doublings_{cell} = 10^7 \ [doublings]$$

$$doublings_{chemostat} \approx \begin{cases} 10^{-5} \ for \ a \ large \ chemostat \ (10^{12} cells, OD_{600} = 1, V_C = 1L) \\ 10^1 \ small \ chemostat \ (10^6 cells, OD_{600} = 0.001, V_C = 0.001L) \end{cases}$$

We reach an important conclusion here. In practice, for a "large" chemostat we expect the time until an advantageous mutation occurs ($T_{mutant}$) to be substantially shorter than the time for an advantageous strain to take over the chemostat population ($T_{takeover}$, as will be addressed in section D), and not necessarily so in a "small" chemostat.

This difference between "large" and "small" chemostats was experimentally analyzed in the work of Egli et. al. [2], where a chemostat of $10^{11}$ cells showed "clock-like" repetitive takeover sweeps and a chemostat of $10^7$ cells showed only random takeover sweeps with long intervals between them. These results also fit with the analysis of [45].

## C5) How many cell doublings and chemostat doublings are needed in the general case for N SNPs and a target size θ?

The number of cell and chemostat doublings required is:

$$doublings_{cell} = \frac{1}{N!} \frac{1}{(\theta \cdot R_{mut})^N} \ [doublings]$$

$$doublings_{chemostat} = \frac{1}{N!} \frac{1}{(\theta \cdot R_{mut})^N \cdot N_{cells}} \ [doublings]$$

In an order of magnitude calculation, for a mutation rate of $R_{mut} \approx 10^r$, a target size of $\theta \approx 10^g$ and a chemostat with a volume and OD corresponding to $N_{cells} \approx 10^n$ we get $doublings_{chemostat} \approx 10^{-Ng-Nr-n}$. For the desired mutation to appear quickly we require the number of chemostat doublings needed on average to be small (<10), or:

$$(-Ng - Nr - n) \leq 0$$

and for r = -10, n = 11 we require the target size to be:

$$g \geq 10 - \frac{11}{N}$$

Where $g$ is the $log_{10}$ of the target size. For N=1 SNP any target size is sufficient. For N=2 SNPs the required target size is on the order of $10^{4-5}$, which is possible, but not common [46]. N>2 SNPs are impossible to achieve in a single doubling of a normal sized chemostat, regardless of the target size.

In summary, for a sufficiently large chemostat, an advantageous mutant strain with a desired (specific) SNP is expected to appear in a short period of time. A strain that needs two SNPs together to become advantageous is possible only when the target sizes are sufficiently large. Other mutations are expected only through successive advantageous takeover sweeps. This can only occur if the SNPs are each individually advantageous. These results show that successive takeovers are the only reliable way (i.e. that does not rely on luck against impossibly low odds) for evolution to proceed in a chemostat.

**C6) How many cell and chemostat doublings, on average, are needed for every single SNP to appear?**

In this question we refer to a seemingly extreme scenario where we require every possible single SNP to exist in at least one cell in the chemostat. It is shown (see appendix C6) that this scenario is actually quite likely, for a sufficiently large chemostat. On the other hand, as was shown above (section C3) a combination of two specific SNPs is almost impossible (let alone every combination of two SNPs). These results can help give intuition to the genetic variability expected inside a chemostat.

A SNP in every possible position within the population:

The number of single cell doublings needed for there to be a SNP in every possible position is (see appendix for derivation):

$$doublings_{cell} \approx \frac{\ln G}{R_{mut}}$$

Using our default characteristic numbers:

$$doublings_{cell} \approx 10^{11} [doublings]$$

Chemostat doublings

As in the cases in the above questions (C2-C5) the number of chemostat doublings needed to achieve the required $doublings_{cell}$, and therefore every single SNP, is:

$$doublings_{chemostat} = \frac{doublings_{cell}}{N_{cells}} = \frac{\ln G}{R_{mut} \cdot N_{cells}}$$

and for *E. coli*:

$$doublings_{chemostat} = \frac{\ln G}{R_{mut}} \cdot \frac{1}{OD_0 \cdot V_C \cdot 10^{12}}$$

Using our default characteristic numbers:

$$doublings_{chemostat} = \frac{10^{11}}{2.5 \cdot 10^{11}} = 0.4$$

Meaning, we can expect every possible one SNP mutant to be present in the chemostat after less than one doubling in our default characteristic case, and certainly in the "large" $10^{12}$ cell chemostat.

For a "small and dilute" chemostat of $10^7$ cells the number of chemostat doublings needed is:

$$doublings_{chemostat} = \frac{10^{11}}{10^7} = 10,000 \; doublings$$

For a doubling time of approx. 10 hours, we get: $Total \; Time \approx 100,000 \; hours \approx 10 \; years$.

Comparing to the required number of doublings required for every single SNP to the number of doublings required for a specific SNP (section C1-C2), we find:

$$doublings_{cell} \; for \; every \; SNP = \ln G \cdot (doublings_{cell} \; for \; specific \; SNP)$$
$$doublings_{chemostat} \; for \; every \; SNP = \ln G \cdot (doublings_{chemostat} \; for \; specific \; SNP)$$

In summary, for a sufficiently large chemostat every possible single SNP is expected to appear, but not every combination of two SNPs. Therefore, strains with more than one SNP are substantially more likely to exist if the first SNP has a selective advantage.

## C7) How do the results vary with different species?

We note that the size of the genome is relevant only in the cases where we require all possible SNPs (section C6) . In the cases where a specific set of advantageous mutations is desired, the genome size surprisingly does not matter (sections C1-C5).

This is because the mutation rate during replication is per base-pair. The specific target base-pair (or base-pairs) in the genome, for which the mutation that gives the cell a selective advantage, has a fixed probability to mutate regardless of the genome size.

Another way of describing the result: On the one hand, the probability of having a random SNP mutation (in any position) rises linearly with the length of the genome. On the other hand, the probability that a SNP is in the "correct" position to increase fitness decreases linearly with the length of the genome. Thus, the total probability of an advantageous mutation occurring is independent of the genome size.

The dependency on species comes from other differences, such as a difference in mutation rate $R_{mut}$, or the population size for a given set of parameters (for example, there can be two similar chemostats of the same size and the same media concentrations, one with *E. coli* and the other with Budding yeast. The former will have more cells than the latter, influencing the mutation time, as seen in the previous sections).

## C8) What is the effect of a hyper-mutating strain?

We saw in section C5 above that the number of chemostat doublings required for a desired mutant strain to appear is $doublings_{chemostat} \approx 10^{-Ng-Nr-n}$ (where the parameters are defined through the mutation rate $R_{mut} \approx 10^r \left[\frac{SNP}{bp*doubling}\right]$, the target size of $\theta \approx 10^g$ [BP], the cell population size of $N_{cells} \approx 10^n$, and the N required SNPs for an advantageous mutation). For the desired mutation to appear quickly we require the number of chemostat doublings needed on average to be small (<10), which can be written as:

$$(-Ng - Nr - n) \leq 0$$

or

$$-(g + \frac{n}{N}) \leq r$$

Where reasonable values are *r=-10, g=1-2, n=8-12*.

In the case of a hyper mutating strain *r* is larger. As we saw above - for a single SNP (N=1), for a sufficiently large chemostat (large *n*) – a larger *r* is not required. For more than one required SNP (N>1), the mutation rate would need to be orders of magnitude larger for the desired mutation to appear quickly, making the benefit of such a hyper mutating strain questionable.

For example, for g=1, n=9, N=3:

$$-\left(1 + \frac{9}{3}\right) = -4 \leq r$$

Meaning, a hyper-mutation rate of at least $10^{-4} \left[\frac{SNP}{bp*doubling}\right]$ instead of $10^{-10}$ (six orders of magnitude difference) would be needed for the desired mutation to appear quickly. This is well above what hyper mutator strains usually show in terms of mutation rates.

A hyper mutating strain might be beneficial in a case where the chemostat must be constrained to be very small or very dilute for technical reasons.

The above finding, together with the insight that takeover sweeps are expected to occur only if the SNPs are each individually advantageous (section C5 above) might also explain why in practice hyper mutating strains do not necessarily always take over the chemostat population.

## D) Single takeover

**D1) What happens from the time a mutant appears until it takes over the reference population?**

The dynamics are illustrated in fig. D1 below, as follows:

(1) The chemostat dynamics (concentrations, etc.) is dominated and set by the properties of the original strain: the original strain grows at a rate dictated by the dilution D, while the advantageous mutant strain grows at a faster rate.

(2) At a relatively late time point, on the timescale derived in the following sections (D3-D6), the mutant strain takes up a substantial fraction of the cells, and is only a few doublings away from complete takeover. At this stage the chemostat dynamics changes as the mutant utilizes more of the limiting substrate, lowering the limiting substrate from the original steady state concentration $S_0^{st}$ to the new steady state concentration $S_1^{st}$.

(3) This continues until the decrease in the limiting substrate in the chemostat causes the mutant strain (which is now the dominant fraction in the population) to grow at a rate of D and the original strain grows at a rate slower than D, on the way to being completely washed out. As shown in many papers (such as [31], [47],[28]), two or more strains competing for the same limiting substrate should not coexist at steady state, and the strain with the higher specific growth rate should eventually prevail. (It should be mentioned that in practice coexistence does occur. This is the result of factors not taken into account in the simplified model discussed here. For example due to factors not analyzed such as growth on walls of the chemostat, the system not being in steady state but with many genetic variants all changing in fraction and accumulating mutations concurrently, some strains excreting byproducts that are the basis for growth of other strains. Etc. All these are beyond the scope of this paper).

Furthermore, note that even though there may be other mutant strains in the chemostat – the competition is of each of them against the dominant strain, and not against each other (assuming that the strains don't directly interact, but rather "interact" by competing for the limiting substrate, whose concentration is in turn dictated by the dominant strain).

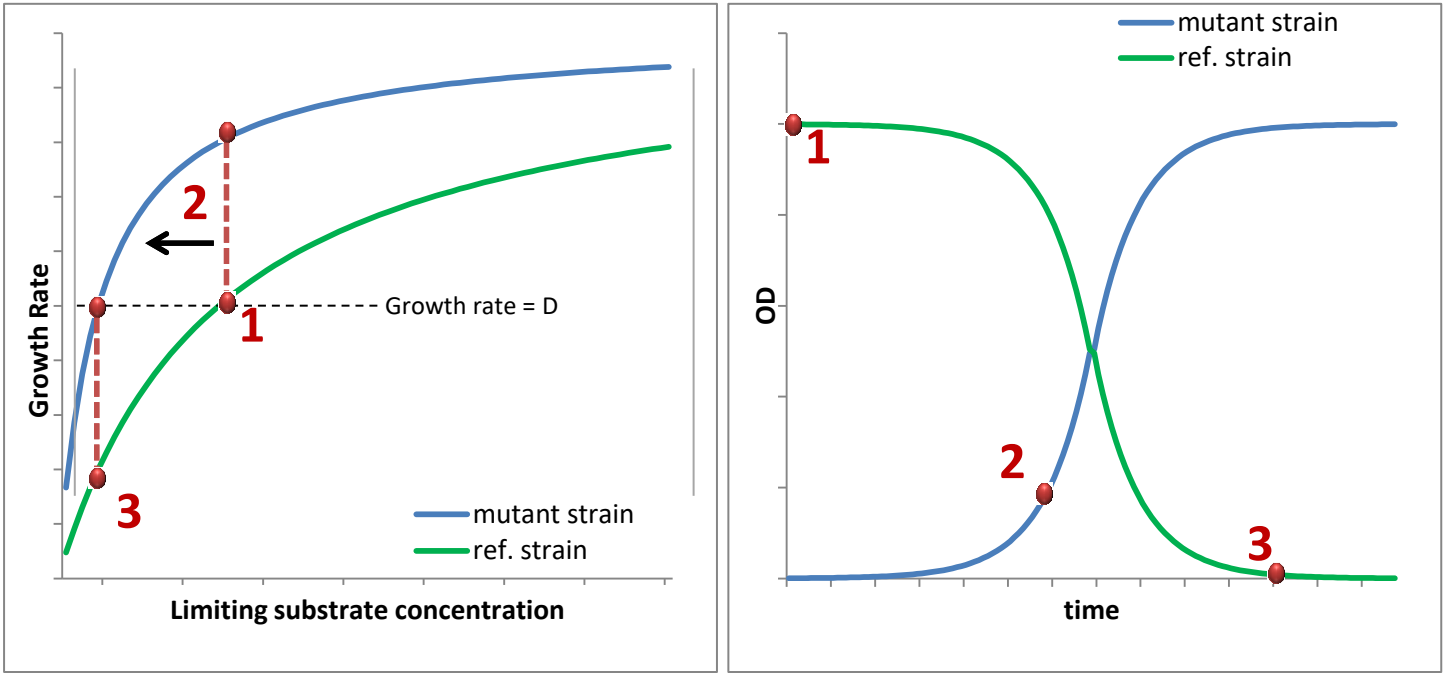

*Fig. D1*: Monod curves(left) and temporal dynamics of OD in the chemostat (right).

**D2) How does an extended version of the Monod relation change the takeover time and dynamics?**

When, instead of the Monod relation $\mu = \mu^{max} \cdot \left(\frac{S}{S+K}\right)$ we use the Korvarova extension (developed to take in to account a non-zero minimal limiting substrate concentration below which there is no growth):

$$\mu = \mu^{max} \cdot \left(\frac{S - S_{min}}{S + K - S_{min}}\right)$$

we get that the growth rates of the original reference strain and mutant strain stay the same as the Monod model (see appendix D2). The only change is that the steady state limiting substrate concentration increases by $S_{min}$. Hence, we regress back to the "classic" problem with no effect on the dynamics or takeover time.

**D3) How long does it take from the time of first appearance of a mutant until takeover?**

When looking at only two competing strains - the takeover time is dependent on D and the cell strain parameters of the original (denoted 0) and mutant (denoted 1) strains – K Monod, maximal and minimal growth rates $(K, \mu^{max}, \mu^{min})$ as follows :

$$T_{takeover} = ln(N_{cells}) \cdot \frac{D + \frac{K_1}{K_0}(\mu_0^{max} - D)}{D^2\left[\frac{K_1 - K_0}{K_0}\right] + D\left[\mu_1^{max} - \frac{K_1}{K_0}(\mu_1^{min} + \mu_0^{max})\right] + \left[\frac{K_1}{K_0} \cdot \mu_1^{min} \cdot \mu_0^{max}\right]}$$

The number of cells required to be considered a takeover is 10% of the total population or: $N_{cells} = OD \cdot (chemostat\ Volume) \cdot 10^{12} \cdot 10$. This formula is somewhat complicated so we will consider special cases where it simplifies in sections D4-D6 below.

**D4) What is the takeover time for a mutant strain with a higher maximal growth rate?**

The takeover time for a mutant strain with a higher maximal growth rate (and otherwise identical parameters $K_0 = K_1$, $\mu_1^{min} = 0$, $\mu_1^{max} > \mu_0^{max}$) is:

$$T_{takeover} = ln(N_{cells}) \cdot \frac{1}{D\left(\frac{\mu_1^{max}}{\mu_0^{max}} - 1\right)} \propto \frac{1}{D}$$

The takeover time monotonically decreases with D, meaning that the faster the dilution the greater the advantage (for a given ratio of growth rates). Since the dilution rate D is a parameter within our control it is advantageous to set D as high as possible to increase selective pressure. The highest practical D equals the maximal growth rate of the reference strain $\mu_0^{max}$, above which the reference strain washes out and selection is impossible.

For our default characteristic values, and $\mu_1^{max} = 0.2$, $\mu_0^{max} = 0.17$ (a growth rate ratio of 1.15 as observed by Egli [2]), the Monod curves and takeover time are shown in Fig. D4:

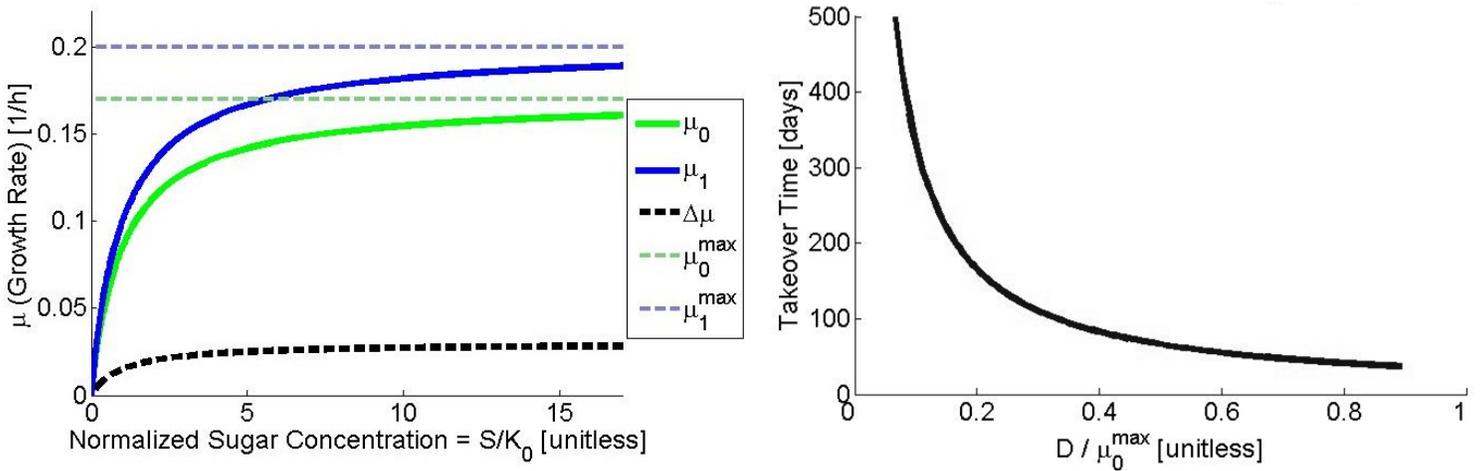

*Fig. D4: Left*: Monod curves for reference strain (green) and mutant strain (blue) with a higher maximal growth rate than the reference strain, with a maximal growth rate ratio of 1.15. Maximum growth rates are denoted by a dashed line in the respective color. Difference between growth rates ( *Δμ*) denoted by black dashed line. *Right*: Takeover time of mutant strain as a function of D. Note that takeover time is proportional to 1/ *Δμ*.

To clarify, since it appears that one should use as fast D as technically possible, it should be mentioned that this is not quite the case. First, using a D faster than the maximal growth rate will lead to washout. Second, a fast D selects specifically for a strain with a higher maximal growth rate. The choice of D for different types of mutant strains is shown in the following sections (D5-D7).

## D5) What is the takeover time for a mutant strain with a smaller K Monod?

The takeover time for a mutant strain with a smaller K Monod - i.e. a strain with higher substrate affinity than the reference strain (and otherwise identical parameters $K_0 > K_1$, $\mu_1^{min} = 0$, $\mu_1^{max} = \mu_0^{max}$) is:

$$T_{takeover} = \ln(N_{cells}) \cdot \left[\left(\frac{K_0}{K_0 - K_1}\right)\frac{1}{(\mu_0^{max} - D)} + \left(\frac{K_1}{K_0 - K_1}\right)\frac{1}{D}\right]$$

This example differs from the case of increased $\mu_1^{max}$ (above, D4) in that the optimal D value - the value at which selection pressure is highest - is an intermediate one. If D is high we will approach the maximal growth rates of both the reference and mutant, which are the same (Figure D5, left). If D is too low, the sugar concentration will approach 0 and both reference and mutant strains will grow at nearly the same (very slow) rate (Figure D5, left). The maximum selective pressure (shortest takeover time) is obtained at:

$$D_{optimal} = \frac{\mu_0^{max}}{\sqrt{\frac{K_0}{K_1}} + 1}$$

For our default characteristic values and $\frac{K_0}{K_1} = 2$ (since it was characteristic of the improvement shown by Egli [2]) the optimal dilution rate is when $\frac{D_{optimal}}{\mu_0^{max}} \approx 0.4$ (in absolute values, $D_{optimal} \approx 0.07 \left[\frac{1}{h}\right]$ or a turnover time of 17 hours) where the takeover time will be approx. 35 days. Note, this value is close to the characteristic working regime of $\frac{D_{optimal}}{\mu_0^{max}} \approx 0.5$. The Monod Curves and takeover time are shown in Fig. D5:

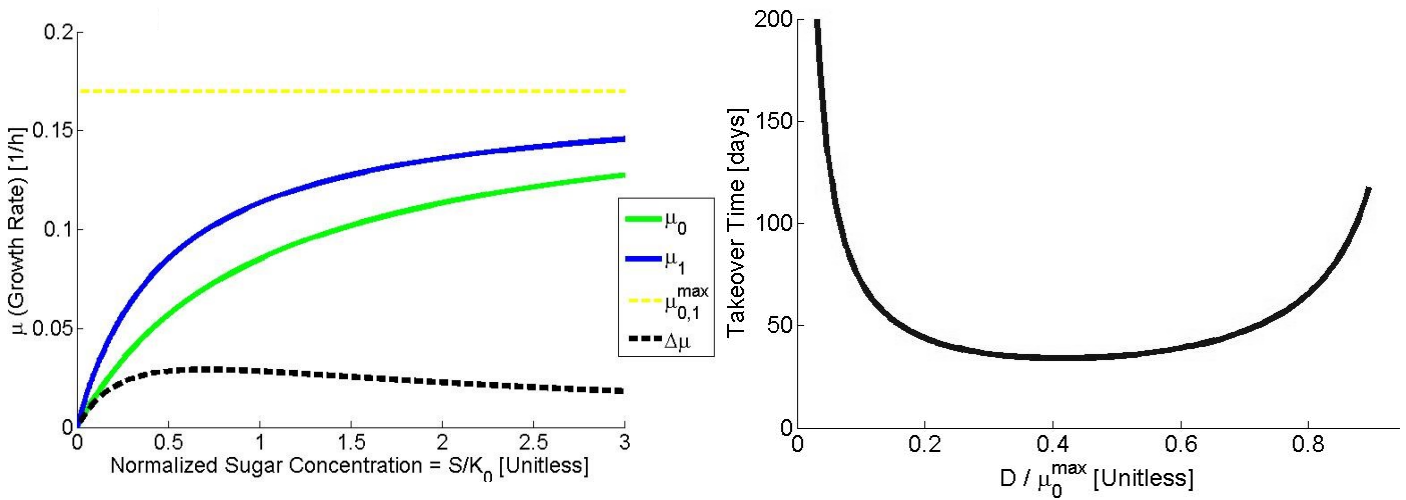

**Fig. D5**: *Left: Monod curves for reference strain (green) and mutant strain with lower k Monod (2\*K₁= K₀) than the reference strain (blue). Maximum growth rate of both strains is denoted by a yellow dashed line. Difference between growth rates (Δμ) denoted by black dashed line. **Right**: Takeover time of mutant strain (blue) as a function of D. Note that takeover time is proportional to 1/Δμ.*

Note, the takeover time is quite robust so long as D is set to a reasonable intermediate value (Figure D5, right). Specifically, the takeover time varies by less than 30% so long as D ranges between $D = 0.15 * \mu_0^{max}$ and $D = 0.75 * \mu_0^{max}$.

**D6) What is the takeover time for a mutant with the ability to grow without the limiting substrate?**

A strain with a non-zero baseline growth rate is a strain which can grow in media with no limiting substrate (presumably by consuming a different substrate present in the media). We assume that the reference strain cannot do this. The takeover time for such a mutant (with otherwise identical parameters $K_0 = K_1$, $\mu_1^{min} > 0$, $\mu_1^{max} = \mu_0^{max}$ ) is:

$$T_{takeover} = ln(N_{cells}) \cdot \frac{1}{\mu_1^{min}} \cdot \frac{1}{\left(1 - \frac{D}{\mu_0^{max}}\right)}$$

The takeover time is monotonically increasing with D, meaning the slower the growth the greater the advantage. This can be rationalized as follows: the mutant grows at a rate $\mu_1^{min}$ even when there is no limiting substrate in the media. However in those conditions the reference strain does not grow at all. Therefore, slowing the dilution has the consequence of slowing the reference strains growth much more than the mutant's. This situation contrasts to results for scenarios of mutants with different $\mu^{max}$ or K Monod (sections D4 and D5 above). In this case, the shortest takeover time is when D=0, the reference strain does not grow, and the mutant strain only needs to multiply from a single cell to $N_{cells}$. Using our default characteristic values, and using $\mu_1^{min} = 0.05 \left[\frac{1}{h}\right]$, equivalent to a doubling time of approx. 14 hours when no limiting substrate is present (where the maximal growth rate is equivalent to 4 hours), the minimal takeover time is around 25 days. The Monod curves and takeover time are shown in Fig. D6:

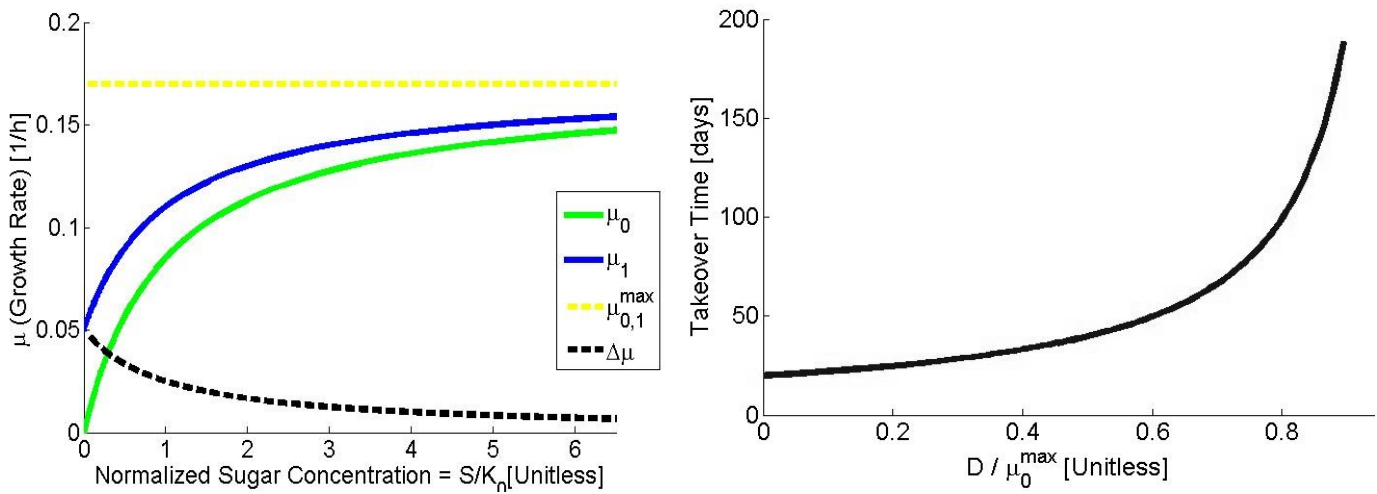

**Fig. D6**: Left: Monod curves for reference strain (green) and mutant strain (blue) with a baseline growth rate ( $\mu_1^{min} > 0$ meaning, there is growth even without any limiting substrate). Maximum growth rate of both strains is denoted by a yellow dashed line. Difference between growth rates (Δμ) denoted by black dashed line. **Right**: Takeover time of mutant strain as a function of D. Note that takeover time is proportional to 1/ Δμ.

**D7) What is the takeover time for a strain with all the above improvements?**

In the previous sections, D4-D6, we examined each of the three improvement types separately. Here we examine a mutant strain with all the types together $K_0 > K_1$, $\mu_1^{min} > 0$, $\mu_1^{max} > \mu_0^{max}$. The equation for takeover time is shown in section D3.

Using our default characteristic values and $\mu_1^{max} = 0.2$, $\mu_0^{max} = 0.17$, $K_0 = 2*K_1$, and $\mu_1^{min} = 0.05$, the Monod curves and takeover time are shown in Fig. D7:

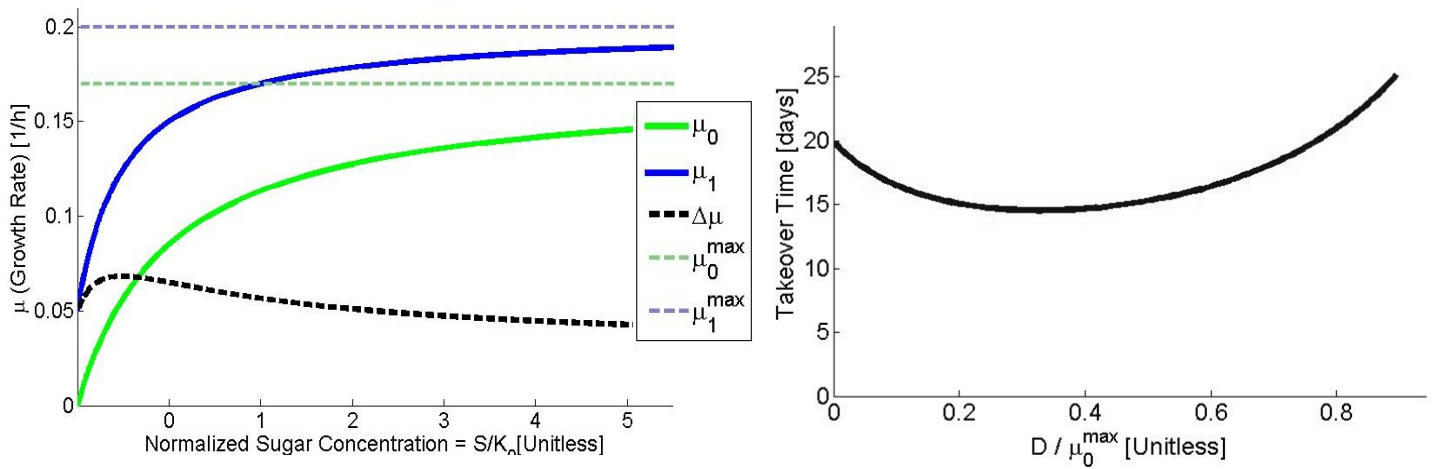

*Fig. D7*: Left: Monod curves for reference strain (green) and mutant strain with a baseline growth rate, higher maximal growth rate, and a smaller K Monod than the reference strain (blue). Maximum growth rates are denoted by a dashed line in the respective color. Difference between growth rates (Δμ) denoted by black dashed line. **Right**: Takeover time of mutant strain as a function of D. Note that takeover time is correlative to 1/ Δμ.

Note, the takeover time is quite robust. This is due to the fact that each improvement gives a selective advantage in a different range of D, so that there is no regime in which the takeover is "ruined" (assuming improvement in all parameters concurrently), as shown in figure D7B below. The takeover time is governed by the need of few mutant cells to multiply to the order of $N_{cells} \approx 10^{11}$ cells, which requires around 25 around 25 generations.

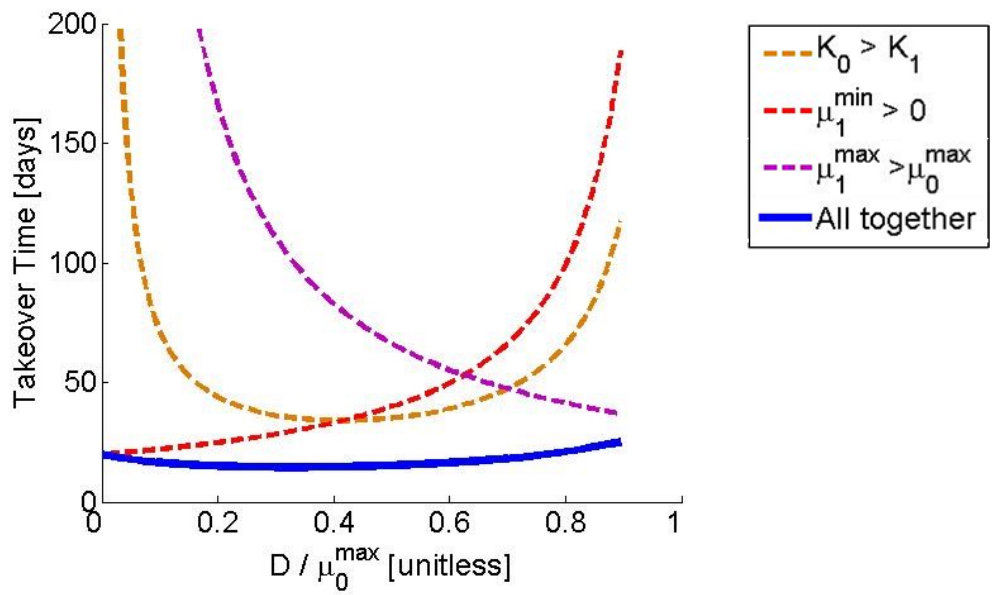

***Fig. D7-B***: *Shows the takeover time of the three types of advantageous mutants separately, and together: [BROWN] lower K Monod, as shown in section D5, [RED] baseline growth, as shown in section D6, [PURPLE] higher growth rate, as shown in section D4, [BLUE] all the improvements together.*

# E) Successive takeovers

In previous chapters we referred to four consecutive stages of the chemostat selection process: (A) parameter choice and setup, (B) basic "steady state" growth, (C) mutation, and (D) single takeover. However, we saw in chapter (C) that a single mutation takeover is usually insufficient to attain a specific desired mutant strain, especially when multiple SNPs are required. Rather, for the full selection process of a desired strain, successive mutation takeover sweeps are needed, where each mutant strain gives a selective advantage on its own.

In this chapter the behavior of successive takeovers is illustrated. This is done using a time-step dynamic computer simulation described in the corresponding appendix E. The influence of two main parameters on the successive takeover behavior is examined: the target size $\theta$ (the number of different SNP locations that give rise to an advantageous mutation), and the improvement in affinity to the limiting substrate $\lambda$.

As discussed in section B2, not all takeovers are detectable on the macro level by monitoring the OD and substrate levels. This chapter shall demonstrate that even when the chemostat seems to be at steady state at the macro level – multiple takeover sweeps can occur. Furthermore, some, but not all, of these takeovers are detectable and differentiable on the microscopic level, using molecular tests like sequencing.

This chapter differs from the previous chapters in that it is not a fifth independent stage in the temporal dynamics of a chemostat, but rather a repetition of stages B-D. Each new strain takes over the chemostat population (B), from it mutates a newer advantageous strain (C) which in turn takes over (D).

### E1) What is the expected behavior of successive takeovers?

The simulation used for this chapter shows the time dynamics of the successive mutant strains. Each new "child" strain mutates from its direct predecessor "parent"; i.e. strain $i=1$ "mutates" out of strain $i=0$, strain $i=2$ from $i=1$ etc.

The children have all the same parameters as their parent, except for a higher "effective" affinity to the limiting substrate, characterized by a smaller K Monod. For our purposes, the biological origin of the selective advantage is immaterial - be it faster transporters, alternative metabolic regulation, improved energy efficiency, etc. The stronger "effective" affinity (smaller K Monod) means that the overall growth rate is faster for the same concentration of limiting substrate.

In this fashion, there are two parameters that define the chain of successively better mutant strains: The first, the population size of strain i required for strain i+1 to appear (related to the target size $\theta$, the number of different SNP locations that give rise to an advantageous mutation). The second, the improvement in affinity to the limiting substrate, which is defined as $\lambda \equiv \frac{K_{i+1}}{K_i}$.

For simplicity, we assumed that both of these parameters stay constant throughout the simulation (i.e. $\lambda \equiv \frac{K_{i+1}}{K_i} = \frac{K_{i+2}}{K_{i+1}}$, etc.).

The way the simulation in defined, takeovers do not go on indefinitely, and a predetermined cell strain is the final one. We chose the "final" takeover to represent a new strain that can grow without a need for the originally limiting substrate (this choice is due to the fact that the simulation was developed as an aid for the study of Antonovsky et al. [1]). The parameters used to represent this final strain are a higher affinity and a higher yield. The affinity - like with the previous improved strains. The higher yield (the mass of cells or product formed per unit mass of substrate consumed) - since the cells utilize another mass source together with the limiting substrate, more biomass can be produced for the same amount of limiting substrate consumed. Note, the previous mutant improvements only increased the **rate** in which biomass is produced, whereas here the **total amount produced** increases.).

Figure E1 shows the results for $\lambda = 1.7$, a new strain arising when the previous strain reaches $10^8$ cells (equivalent to a target size of $\theta \approx 10^3$ [bp]), and the "final strain" appears at i=6.

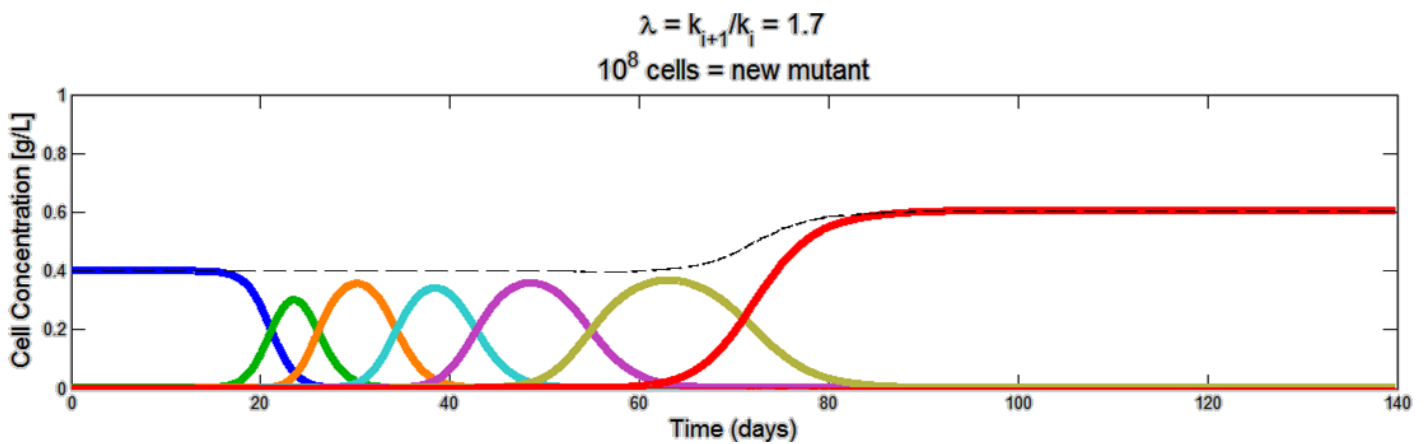

**Fig. E1**: Time dynamics of successive takeovers, for $\lambda = 1.7$, a new strain arising when the previous strain reaches $10^8$ cells (equivalent to a target size of $\theta \approx 10^3 [bp]$), and the "final strain" appears at i=6. The total cell concentration is $X_0^{st} = 0.4 \left[\frac{g}{L}\right]$ (approx. OD = 1). The influx feed is $S_d = 0.1 \left[\frac{g}{L}\right]$.

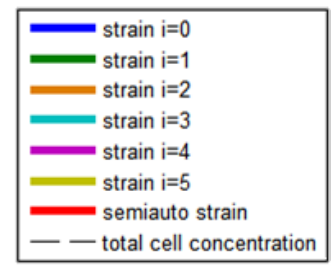

A few observations are worth noting, as follows.

Even though the mutations are similar, the takeover temporal "peaks" are not of uniform height, width, nor exactly evenly spaced in time. This is because the background population and residual limiting substrate concentration change over time.

After the first takeover, which takes a long time to emerge, the subsequent strains appear one after the other relatively fast. The difference between the first takeover and the rest is that strain i=1 has to overtake a population of ~$10^{11}$-$10^{12}$ cells of strain i=0, whereas strain i=2 first appears when i=1 consists of only ~$10^8$ cells and most of the cells are still of i=0 and thus the fitness advantage of i=2 over the background cells enjoys the benefits of both the first and second mutation when it grows in parallel to i=1 taking over i=0. A similar scenario happens with the subsequent strains.

Once the mutant strains start taking over the population, they do so in a relatively timed and paced manner (although not exactly). This behavior was dubbed "clock-like" by Egli [2].

Finally, the total cell concentration (equivalent to OD, denoted by the dashed black line) stays almost constant, until the final strain takes over. Together with the fact that the limiting substrate is at undetectably low concentrations – the successive takeovers are undetectable without using molecular tests like sequencing. For example, when in the example above strain i=2 reaches its peak (at about day 30) there are at least four strains in the chemostat in substantial quantities, i = 0,1,2,3. Yet, they still add up exactly to the original steady state concentration, due to conservation of the mass originating in the limiting substrate influx (see B2 for elaboration).

**E2) How does the value of the improved affinity to the limiting substrate, $\lambda$, impact the successive takeover behavior?**

In the following figure (E2) are the results of the simulation for a new strain arising when the previous strain reaches $10^7$ cells (equivalent to a target size of $\theta \approx 10^4$ [bp]), a "final strain" at i=6, and three different values of the improved affinity to the limiting substrate, $\lambda = 1.2, 1.5, 1.7$

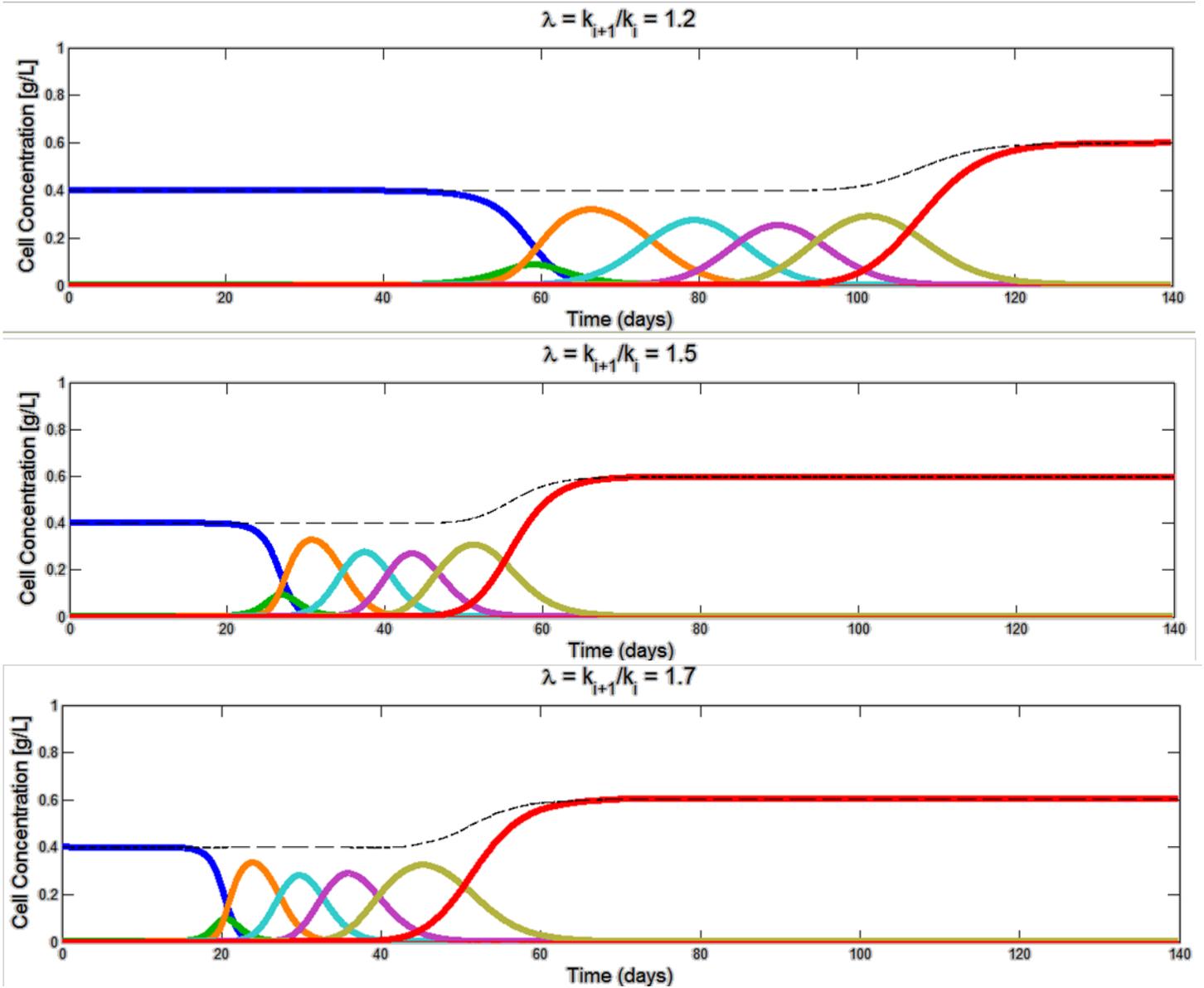

**Fig. E2**: Time dynamics of successive takeovers, for $\lambda = 1.2, 1.5$ and $1.7$, a new strain arising when the previous strain reaches $10^7$ cells (equivalent to a target size of $\theta \approx 10^4$ [bp]), and the "final strain" appears at i=6. The total cell concentration is $X_0^{st} = 0.4 \left[\frac{g}{L}\right]$ (approx. OD = 1). The influx feed is $S_d = 0.1 \left[\frac{g}{L}\right]$.

The behavior of all three options is similar (relative peak dimensions and timing), where the difference between them is a compression of the time scale. As expected, a larger affinity ratio $\lambda$ results in a shorter overall takeover time.

**E3) How does the value of the target size $\theta$, or corresponding population size required for a new mutant strain to arise, impact the successive takeover behavior?**

In the following figure (E3) are the results of the simulation for an improved affinity to the limiting substrate $\lambda=1.7$, a "final strain" at $i=6$, and a new strain arising when the previous strain reaches $10^6$, $10^7$, $10^8$ or $9*10^9$ cells (equivalent to a target size of $\theta \approx 10^5$, $10^4$, $10^3$ or 10 [bp], respectively).

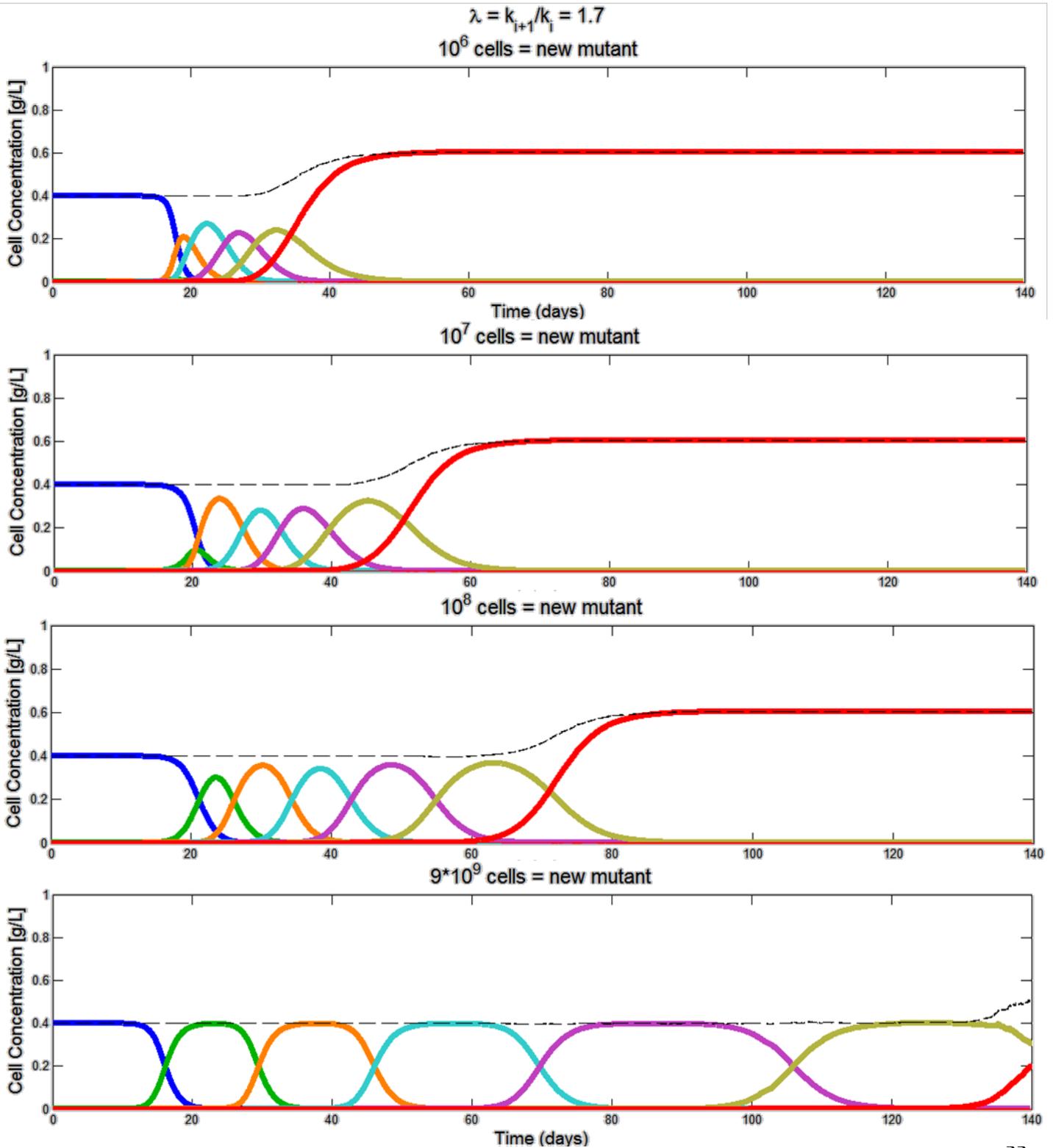

**Fig. E3**: Time dynamics of successive takeovers, for λ = 1.7, a new strain arising when the previous strain reaches $10^7 - 9*10^9$ cells (equivalent to target sizes of $\theta \approx 10^{5-1}$ [bp] respectively), and the "final strain" appears at i=6. The total cell concentration is $X_0^{st} = 0.4 \left[\frac{g}{L}\right]$ (approx. OD = 1). The influx feed is $S_d = 0.1 \left[\frac{g}{L}\right]$.

As expected, a larger population size requirement (smaller target size) results in a longer overall takeover time. However, the behavior here differs between cases more than it did in the previous question: the peaks change in height, width and relative spacing.

**E4) Is it always possible to detect successive takeovers? Might they appear as a single cohort?**

There are cases where a set of successive takeovers happen rapidly enough so that it is very difficult to differentiate between them, even when using DNA sequencing to determine allele frequency in the chemostat population. Thus, multiple successive takeovers of strains (each with a single advantageous mutation) would falsely appear as a single takeover of a strain with multiple mutations, or a "cohort strain".

The following set of figures (E4) illustrates such a phenomenon. Figure E4 consists of two pairs of panels. The first pair (E4A, E4B) shows a case where the successive takeovers can be differentiated one from another using sequencing. The second pair (E4C, E4D) shows a case where such differentiation is very difficult, and the successive takeovers appear as cohort strain. In each pair, the first figure (E4A, E4C) shows the time dynamics of strain cell concentrations as above. The second figure (E4B, E4D) shows the time dynamics of an equivalent of the "allele frequency" of each mutant strain, in that we assume that every strain i has some allele signature that appears in that strain and in all subsequent strains > i. Since the simulation models only a single lineage, all alleles eventually reach a frequency of 1 (100% of the population).

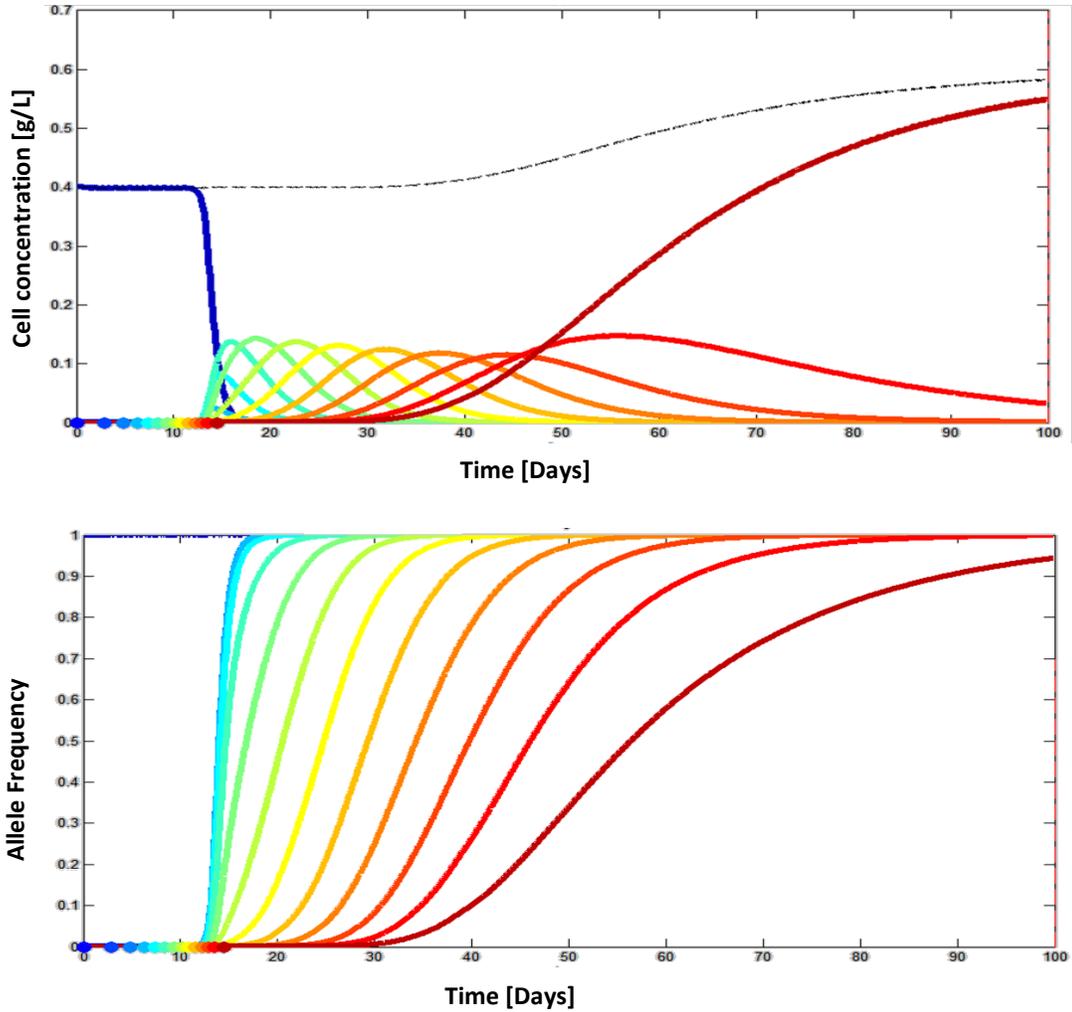

**Fig. E3:** [**A (top)**: Time dynamics of cell concentration during successive takeovers. **B (bottom):** Time dependent allele frequency]. $\lambda = 1.4$, a new strain arising when the previous strain reaches $10^{3.5}$ cells (equivalent to a target size of $\theta \approx 10^{7.5}$ [$bp$]), and the "final strain" appears at i=15. The total cell concentration is $X_0^{st} = 0.4 \left[\frac{g}{L}\right]$ (approx. OD = 1). The influx feed is $S_d = 0.1 \left[\frac{g}{L}\right]$. the strains are rainbow-colored by order, starting with the first strain i=0 in blue, and ending with strain i=15 in red . The time of the first appearance of each strain is denoted on the X axis with a dot of the same color.

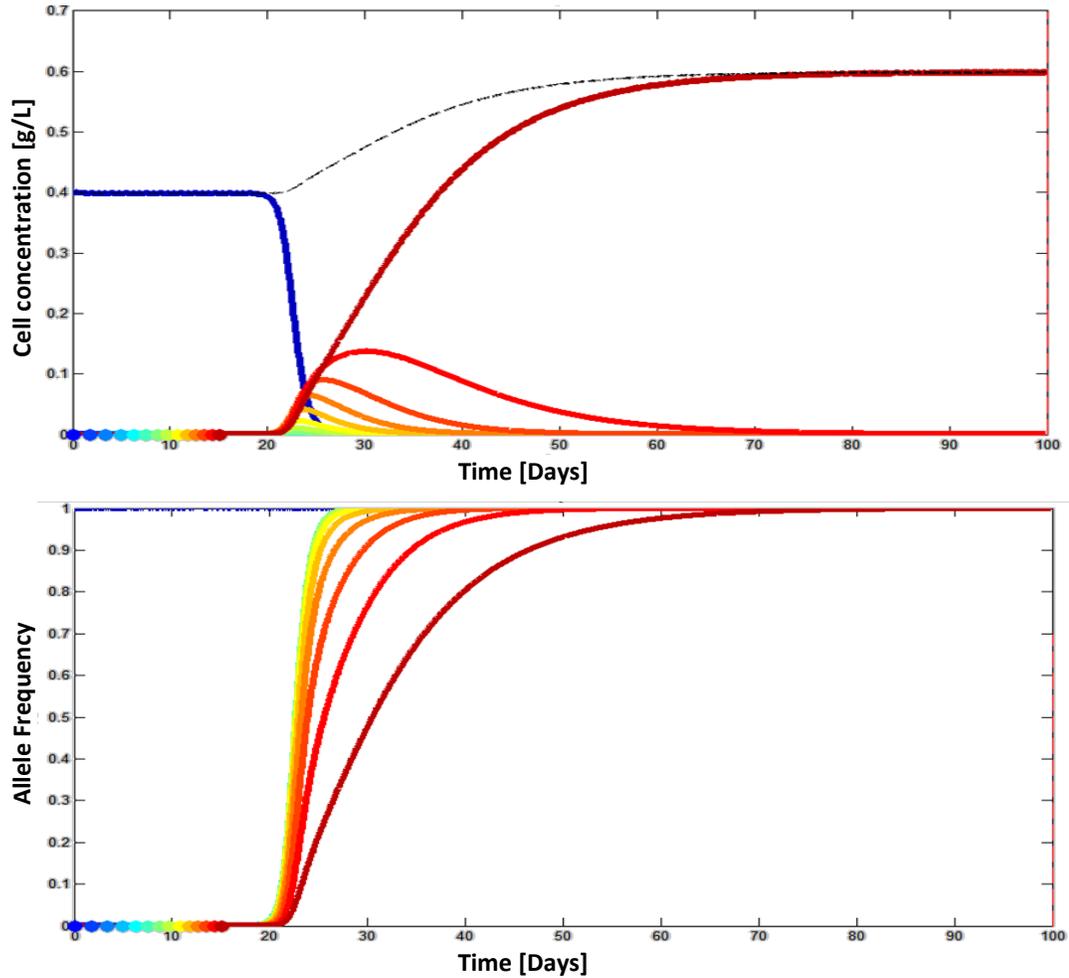

**Fig. E3** [**C (top)**: Time dynamics of cell concentration during successive takeovers. **D (bottom):** Time dependent allele frequency]. All parameters as above (E4A-B), except $\lambda = 1.1$, a new strain arising when the previous strain reaches $10^3$ cells (equivalent to a target size of $\theta \approx 10^8$ [$bp$]). "Cohort formation" is visible.

The first case (E4A, E4B) and the second case (E4C, E4D) have almost all the same parameters: a total cell concentration of $X_0^{st} = 0.4 \left[\frac{g}{L}\right]$ (approx. OD = 1), and limiting substrate influx feed of $S_d = 0.1 \left[\frac{g}{L}\right]$, and a "final strain" at i=15. The two cases differ only in improved affinity ratio (a moderate improvement of $\lambda = \frac{K_{i+1}}{K_i} = 1.4$ for the first case and a small improvement $\lambda = 1.1$ for the second), and in the population size required for a new strain to appear ($10^{3.5}$ cells for the first and $10^3$ for the second, equivalent to large target sizes of $\theta \approx 10^{7.5}$ and $10^8$ [bp], respectively).

In all the figures the strains are rainbow-colored by order, starting with the first strain i=0 in blue, and ending with strain i=15 in red. The time of the first appearance of each strain is denoted on the X axis with a dot of the same color.

In the first case (E4A-B), a few findings are apparent.

For the chosen parameters, each mutant strain first appears in the chemostat a long time before it reaches a detectable fraction of the population.

Due to the large target size, the strain takeovers occur with a background dominant population made up of distant, rather than immediate, ancestors of the growing mutant strain; because the immediate ancestor has not yet had enough time to take over the population. Since only strains that take up a substantial portion of the population influence the limiting substrate concentration, the mutant strains compete against those older strains. The relevant affinity improvement is therefore higher than $\lambda$, (specifically, a power of $\lambda$, depending on the number of strains between the dominant strains and the mutant strain).

The takeover rate diminishes the more "advanced" the strain, even though the improvement ratio $\lambda$ stays constant. This is due to the compound nature of the improvements.

The second case (E4C-D) shows an example of "cohort appearance". A periodic sequencing to determine allele frequency, for instance at days 0, 5, 10, 15, 20, 30, would likely lead to the false conclusion that a single strain with 12 mutations (at least) took over the population in one sweep, wherein actually 12 sequential mutant strains arose one from another. Furthermore, one might conclude that not all of the 12 mutations were beneficial, or that only some combination of mutations gives a selective advantage, when actually each one of the 12 was advantageous on its own, relative to the strains' ancestor.

The cohort appearance is due to the combination of large target size, leading to many mutant strains, and small affinity improvement, leading to mutant strains not gaining enough of an advantage before the subsequent strain appears.

This finding can help explain experimental cases that have previously been only partially deciphered. For example, Desai et. al. [48] study mutation cohorts in the growth of yeast for 1,000 generations. The study finds "Multiple mutations or cohorts of mutations are often present simultaneously… and selective sweeps are often 'nested'; that is, one sweep initiates before the preceding sweep has completed". Several examples of the phenomenon are presented, including cases where each of the individual mutations is proven to be beneficial independently, and repeatedly arise in reproduction experiments. Yet, only a limited explanation is presented, in that "Cohorts … are forms of genetic hitchhiking, in which individual mutations are helped (or hindered) by the genetic background in which they happen to arise. This includes both hitchhiking… as well as 'quasi-hitchhiking' of multiple beneficial mutations that act together as co-drivers". The study concludes: "Further work is needed to determine the mechanism underlying the formation of these cohorts".

Interestingly, in an earlier study Desai and Fisher [49] state that multiple mutations can occur in the same lineage before the first beneficial mutation fixes, where each mutation is individually beneficial, and the strain benefits from the combined mutations. This is a similar to some of our assumptions above, yet the hypothesis of multiple rapid successive takeovers falsely appearing as a "cohort" is not proposed in the later study.

# IV. Appendices with full derivations:

In this section we shall present the full calculations and considerations that resulted in the short answers brought in the "Results - Q&A" section for the relevant questions. Questions that were addressed fully with a short answer are skipped in this section.

## A) Parameter choice and setup

### A2) What is the steady-state concentration of the limiting substrate within the chemostat?

<u>The substrate levels in the chemostat, s, as a function of D</u>
For this stage, we shall assume that the dynamics of the chemostat are determined by the reference strain, and that the system has reached a steady-state regarding the reference strain. Specifically, we assume that s reaches a steady state, which we shall calculate in this section.

The system has reached a steady-state, meaning that the state of the system is only a function of the time within the interval (i.e. everything behaves periodically with a period $\tau_d$). To avoid confusion, we clarify that time $t = 0$ is the time right after the dilution and time $\tau_d$ is immediately before the dilution step.

We can relate between $X(0)$ and $X(\tau_d)$ by observing that each dilution step, we remove a volume of $V_d$ with cells at concentration $X(\tau_d)$ and add the same volume with no cells to get a tank with a concentration $X(0)$. So, from the conservation of mass:

$$V_C \cdot X(\tau_d) = V_C \cdot X(0) + V_d \cdot X(\tau_d)$$

Therefore, $\frac{X(0)}{X(\tau_d)} = 1 - \frac{V_d}{V_t} = \phi$. The average growth rate is defined as

$$\tilde{\mu} \equiv \frac{1}{\tau_d} \ln\left(\frac{X(0)}{X(\tau_d)}\right)$$

So we get

$$\tilde{\mu} = -\frac{\ln(\phi)}{\tau_d}$$

Now, we can write a similar conservation law for the limiting resource. Note that unlike in the case of the cells, the added volume $V_d$ contains a concentration $s_d$ of the limiting resource:

$$V_t \cdot S(\tau_d) + V_d \cdot S_d = V_t \cdot S(0) + V_d \cdot S(\tau_d)$$

Solving for $S(\tau_d)$ we get:

$$S(\tau_d) = \frac{S(0) - (1-\phi) \cdot S_d}{\phi}$$

The third conservation rule we shall use is that the total amount of limiting element in the tank does not change within the time interval between dilutions:

$$\frac{X_0(t)}{Y_0} + S(t) = const.$$

Furthermore, in the dilution step, the total amount inserted equals the total amount removed

$$S_d \cdot V_d = \left(\frac{X_0(\tau_d)}{Y_0} + S(\tau_d)\right) \cdot V_d$$

Therefore:

$$\frac{X_0(t)}{Y_0} + S(t) = S_d$$

$$S(t) = S_d - \frac{X_0(t)}{Y_0}$$

Moreover, the concentration of cells is maximal when $S(t) = 0$. We can define the maximal cell concentration $X_0^{max} = S_d \cdot Y_0$. This would be the cell concentration if all the limiting resource were depleted from the tank (technically an unachievable situation due to the Monod relation).

Using these results, we shall now calculate the time-dependent concentration $X_0$ between the dilution intervals:

$$\mu_0(S(t)) = \frac{\dot{X}_0}{X_0} = \mu_0^{max} \cdot \left(\frac{S(t)}{S(t) + K_0}\right)$$

$$\mu_0(t) = \frac{\dot{X}_0(t)}{X_0(t)} = \mu_0^{max} \cdot \left(\frac{\frac{X_0(t)}{S_d \cdot Y_0} - 1}{\frac{X_0(t)}{S_d \cdot Y_0} - \left(1 + \frac{K_0}{S_d}\right)}\right)$$

Substituting for $X_0^{max}$

$$\frac{\dot{X}_0(t)}{X_0(t)} = \mu_0^{max} \cdot \left(\frac{\frac{X_0(t)}{X_0^{max}} - 1}{\frac{X_0(t)}{X_0^{max}} - \left(1 + \frac{K_0}{S_d}\right)}\right)$$

So, we get to the following differential equation:

$$\frac{d\left(\frac{X_0(t)}{X_0^{max}}\right)}{d(\mu_0^{max} \cdot t)} = \left(\frac{X_0(t)}{X_0^{max}}\right) \cdot \left(\frac{X_0(t)}{X_0^{max}} - 1\right) \cdot \left(\frac{X_0(t)}{X_0^{max}} - \left(1 + \frac{K_0}{S_d}\right)\right)^{-1}$$

We will use the following ODE solution for $\frac{d(Z)}{dt} = (Z) \cdot (Z + a) \cdot (Z + b)^{-1}$:

$$t_{final} - t_{initial} = \frac{b}{a} \ln\left(\frac{Z_{final}}{Z_{initial}}\right) + \left(1 - \frac{b}{a}\right) \ln\left(\frac{Z_{final} + a}{Z_{initial} + a}\right)$$

For:

$$Z \equiv \frac{X_0(t)}{X_0^{max}}, a \equiv -1, b \equiv -\left(1 + \frac{K_0}{S_d}\right)$$

We get:

$$\mu_0^{max} \cdot t = \left(1 + \frac{K_0}{S_d}\right) \ln\left(\frac{X_0(t)}{X_0(0)}\right) - \left(\frac{K_0}{S_d}\right) \ln\left(\frac{X_0(t) - X_0^{max}}{X_0(0) - X_0^{max}}\right)$$

It is difficult to find an analytic expression for $X_0(t)$, but what we can do is find $X_0(\tau_d)$ by applying the boundary condition $X_0(\tau_d) \cdot \phi = X_0(0)$:

$$\mu_1^{max} \cdot \tau_d = -\ln(\phi) - \left(\frac{K_0}{S_d}\right) \ln(\phi) - \left(\frac{K_0}{S_d}\right) \ln\left(\frac{1}{\phi} + \frac{\frac{1}{\phi} - 1}{\frac{X_0(0)}{X_0^{max}} - 1}\right)$$

$$= -\ln(\phi) - \left(\frac{K_0}{S_d}\right) \ln\left(1 + \frac{1 - \phi}{\frac{X_0^{max}}{X_0(0)} - 1}\right)$$

Using the relation $-\frac{\ln(\phi)}{\tilde{\mu}} = \tau_d$:

$$-\frac{\ln(\phi)}{\tilde{\mu}} \mu_0^{max} = -\ln(\phi) - \left(\frac{K_0}{S_d}\right) \ln\left(1 + \frac{1 - \phi}{\frac{X_0^{max}}{X_0(0)} - 1}\right)$$

Solving for $\frac{X_0(0)}{X_0^{max}}$

$$\frac{X_0(0)}{X_0^{max}} = 1 - \frac{1 - \phi}{1 - \phi^{\frac{S_d}{K_0}\left(\frac{\mu_0^{max}}{\tilde{\mu}} - 1\right)}} = 1 - \frac{1 - \phi}{1 - \phi^{-\frac{S_d}{K_0}\left(\frac{\mu_0^{max} \cdot \tau_d}{\ln(\phi)} + 1\right)}}$$

Now we shall calculate the concentration $S(0)$ and $S(\tau_d)$:

$$\frac{X_0(t)}{X_0^{max}} = \frac{Y_0 \cdot (S_d - S(t))}{Y_0 \cdot S_d} = 1 - \frac{S(t)}{S_d}$$

$$S(0) = S_d\left(1 - \frac{X_0(0)}{X_0^{max}}\right) = \frac{(1-\phi)S_d}{1 - \phi^{-\frac{S_d}{K_0}\left(\frac{\mu_0^{max} \cdot \tau_d}{\ln(\phi)} + 1\right)}}$$

Using the above definition: $D = \frac{1-\phi}{\tau_d}$, the substrate right before and right after the dilution is:

$$S(0) = \frac{(1-\phi) \cdot S_d}{1 - \phi^{-\frac{S_d}{K_0}\left(\frac{\mu_0^{max}(1-\phi)}{\ln(\phi)D} + 1\right)}}$$

$$S(\tau_d) = \frac{S(0) + (1-\phi) \cdot S_d}{\phi}$$

Assuming a quasai-continuous chemostat, or $\phi \to 1$:
$$S(\tau_d) \underset{\phi \to 1}{\approx} S(0)$$

$K_0$ is used, because the reference strain heavily dominates the chemostat, and the final $S$ level is determined by it.

Since the substrate levels vary only slightly let us refer to substrate levels as constant within the chemostat, and define the concentration $S_0^{st} \equiv S(0)$, or more generally
$$S_i^{st} \equiv S(0)_{when\ the\ chemo\ is\ dominated\ by\ strain\ i}$$

Let us now simplify $S_0^{st}$:

$$S_0^{st} \equiv S(0) = \frac{(1-\phi) \cdot S_d}{1 - \phi^{-\frac{S_d}{K_0}\left(\frac{\mu_0^{max}(1-\phi)}{\ln(\phi)D} + 1\right)}}$$

$$1 - \phi^{-\frac{S_d}{K_0}\left(\frac{\mu_0^{max}(1-\phi)}{\ln(\phi)D} + 1\right)} = 1 - e^{-\frac{S_d}{K_0}\left(\frac{\mu_0^{max}(1-\phi)}{\ln(\phi)D} + 1\right)\ln(\phi)}$$

For $\phi \sim 1 \to$ Taylor expansion

$$1 - e^{-\frac{S_d}{K_0}\left(\frac{\mu_0^{max}(1-\phi)}{\ln(\phi)D} + 1\right)\ln(\phi)} \approx \frac{-S_d(\phi - 1) \cdot (D - \mu_0^{max})}{K_0 D}$$

Therefore, we can approximate the substrate levels in the chemostat, as a function of D, as:

$$S_0^{st}(D) \approx \frac{K_0 D}{(\mu_0^{max} - D)}$$

Let us note that $S_0^{st}$ is independent of $S_d$. This means that a higher input concentration does not translate into a higher steady-state residual substrate concentration, but rather into a higher $X$ (or a higher OD) (see section A3). Moreover, since $K_0$ and $\mu_0^{max}$ are parameters of the cell strain, the only system parameter that $S_0^{st}$ is a function of is D.

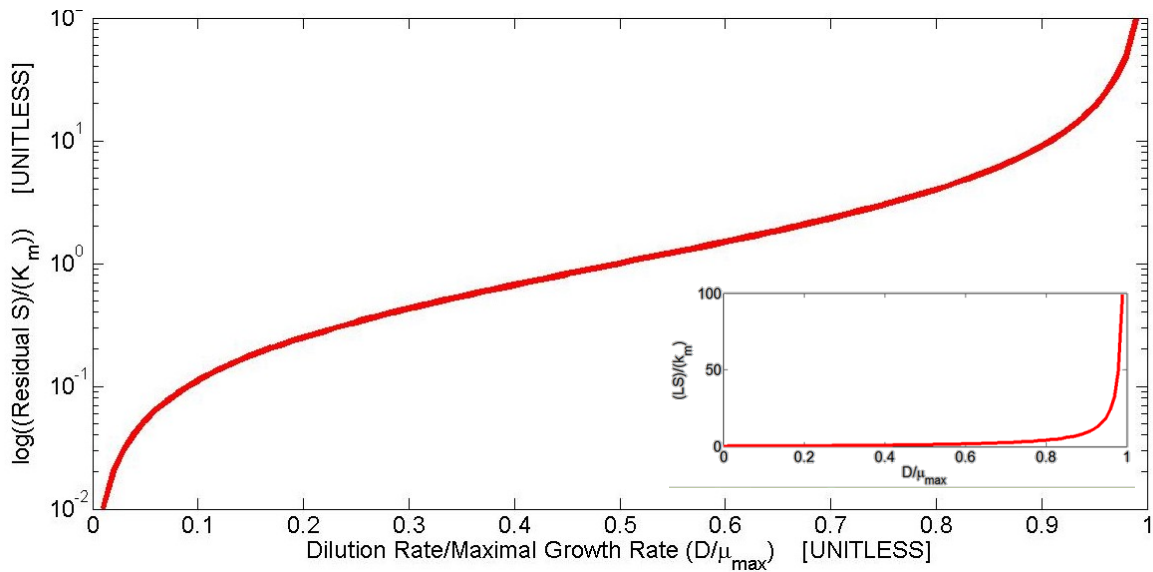

*Fig. B3:* Normalized Residual limiting substrate (s) [log scale] vs. normalized dilution rate (D). Inset figure- linear scale.

## A3) What are the steady state cell concentration and OD?

Using mass conservation (continuing from section A2 above):

$$S_d \cdot V_d = \left(\frac{X_0(\tau_d)}{Y_0} + S(\tau_d)\right) \cdot V_d$$

Using our definition that the steady state concentrations are immediately before an influx drop.

$$S_d = \left(\frac{X_0^{st}}{Y_0} + S_0^{st}\right)$$

Using the previous questions result (A2)

$$X_0^{st} = \left(S_d - \frac{K_0 \cdot D}{(\mu^{max} - D)}\right) \cdot Y_0$$

For *E. coli*, we use $1 \cdot [OD_{600}] = 0.4 \left[\frac{g}{L} \; CDW\right]$ [*taking the average of BNIDS 109838, 107924, 109835, 109837*]:

$$OD_0^{st} = \left(S_d - \frac{K_0 \cdot D}{(\mu^{max} - D)}\right) \cdot \frac{Y_0}{0.4}.$$

For our default characteristic values:

$$X_0^{st} = \left(0.1 - \frac{10^{-3} \cdot 0.07}{(0.17 - 0.07)}\right) \cdot 2 = X_0 - 10^{-3} \approx 0.199 \; \left[\frac{g}{L}\right]$$

$$OD_0 = 0.497$$

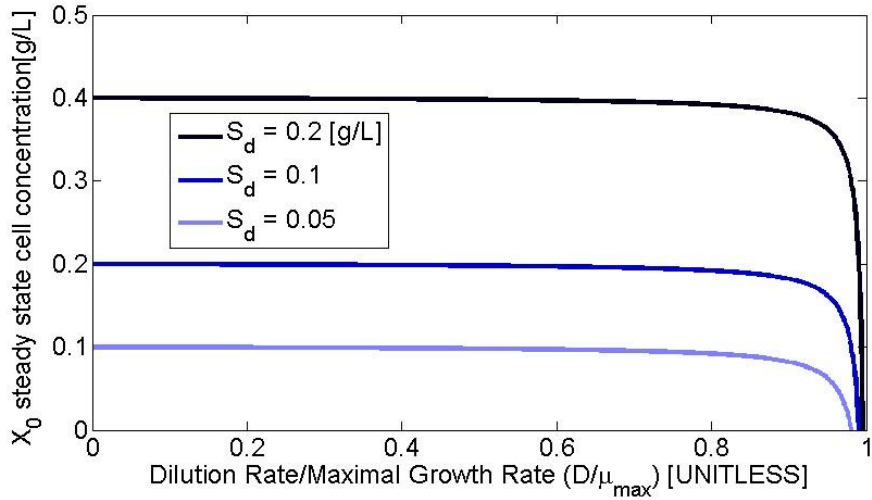

***Fig. A3****: The steady state cell concentration, as a Function of D, for different the input limiting substrate concentrations ($S_d$). The cell concentration is almost constant, except for D values close to the maximal growth rate, where washout is approached.*

For comparison, for *S. cerevisiae* (Budding yeast) with similar strain growth parameters ($Y_0, K_0, \mu^{max}$), we use $1 \cdot [OD_{600}] = 0.6 \left[\frac{g}{L} CDW\right]$ [*assuming similarity to BNID 111182*]:

$$OD_0^{st} = \left(S_d - \frac{K_0 \cdot D}{(\mu^{max} - D)}\right) \cdot \frac{Y_0}{0.6} = 0.3 \; \boldsymbol{Budding\; yeast}$$

Let us note that the steady state cell concentration is dependent on the input limiting substrate concentration ($S_d$), unlike the steady state limiting substrate $S_0^{st}$(see A2 above). This means that a higher input concentration translates into a higher steady-state cell concentration. Furthermore, since $S_0^{st}$ is dependent on only one system parameter ($D$) and $X_0^{st}$ is dependent on two system parameters ($D, S_d$), $S_0^{st}$ and $X_0^{st}$ can be independently set.

**A4) What is the number of cells in the chemostat at steady state?**
The cell concentration for an *E. coli* culture with $OD^{600}$ of 1 is *[BNID100985]* (assuming the relation between OD and cell number is in a linear regime):

$$1 \cdot [OD^{600}] \approx 10^{12} \frac{cells}{L}$$

The $OD_{600}$ of the chemostat is (using the relation from section A3 above):

$$1 \cdot [OD^{600}] = 0.4 \left[\frac{g}{L} CDW\right] \rightarrow$$
$$OD_0 = \left(S_d - \frac{K_0 \cdot D}{(\mu^{max} - D)}\right) \cdot \frac{Y_0}{0.4}$$

So the number of cells in the chemostat is:

$$N_{cells} \approx OD_0 \cdot V_C \cdot 10^{12} = \left(S_d - \frac{K_0 \cdot D}{(\mu^{max} - D)}\right) \cdot \frac{Y_0}{0.4} \cdot V_C \cdot 10^{12}$$

And for our default characteristic values:

$$N_{cells} \approx 3 \cdot 10^{11}$$

For comparison, the cell concentration for a Budding Yeast culture with $OD^{600}$ of 1 is *[BNID100986]*:

$$1 \cdot [OD^{600}] = 3 \cdot 10^{11} \frac{cells}{L}$$

$$N_{cells} = OD_0 \cdot V_C \cdot 3 \cdot 10^{11} = \left(S_d - \frac{K_0 \cdot D}{(\mu^{max} - D)}\right) \cdot \frac{Y_0}{0.6} \cdot V_C \cdot 3 \cdot 10^{11}$$

Assuming similar growth parameters ($Y_0, K_0, \mu^{max}$) as *E. coli*:

$$N_{cells}^{Budding\ yeast} = 5 \cdot 10^{10}$$

## A5) What is the average cell growth rate and doubling time at steady state? When should a factor of ln(2) be used?

In the corresponding section A5 above, the doubling time is $\frac{\ln(2)}{D}$, which is smaller by a multiplicative factor of ln(2), about 0.7, from the chemostat turnover time $\frac{1}{D}$. Here we give an intuitive explanation for that difference by examining the hypothetical continued growth of the cells that have already been removed from the chemostat during dilution, or "waste cells", in the time after they had been removed.

Going back to the definition of the dilution rate D, it equals "the fraction of the chemostat volume replaced per hour by inflow from the reservoir". Because the system is at steady-state - dilution and growth must balance exactly to have no net change in the cell concentration. Since the concentration for cells is constant, we conclude that over 1/D hours $V_c * X_0^{st}$ cells were removed.

Thus, on the one hand, naively, since at the start there are $V_c * X_0^{st}$ cells and **after 1/D hours** it seems like there is a total of $2 * V_c * X_0^{st}$ cells - the growth that occurred in this time frame should thus be equivalent to one doubling of the cell population. But, on the other hand, if we calculate directly from the growth equations the time required for one doubling of the cell population - we find the time required to be shorter, $\frac{\ln(2)}{D}$, as follows:

Growth rate is defined as: $\dot{X} = \mu \cdot X$. The solution is:

$$X = X_{t=0} \cdot e^{\mu t}$$

Since $\mu = D$, $X = X_{t=0} \cdot e^{Dt}$

A doubling occurs after:

$$X(t = 0) = X_{t=0}$$

$$X(t^{doubling}) = X_{t=0} \cdot e^{D*t^{doubling}}$$

$$2 * X_{t=0} = X_{t=0} \cdot e^{D*t^{doubling}}$$

$$t^{doubling} = \frac{\ln(2)}{D}$$

In other words, if the population doubled after $\frac{\ln(2)}{D}$ hours, but on the other hand after 1/D hours there are still only $2 * V_c * X_0^{st}$ cells – where are the "missing cells" that should have grown in the time difference $\frac{1-\ln(2)}{D}$?

As mentioned, the cells that have been removed from the chemostat, or "waste cells", in principle continue to grow. We shall now prove that the extra growth exactly makes up for the "missing cells".

At each dilution pulse, at a time interval of $\tau_d$, a "batch" of $V_d * X_0^{st}$ cells is removed from the chemostat. The first batch has D hours remaining to grow until the chemostat turnover is complete. If those cells were to continue growing at the same rate $\mu = D$, then at the final time they would total:

$$(Cell\ total\ at\ chemostat\ turnover\ time\ D)_{first\ batch} = V_d * X_0^{st} * e^{D/D}$$

The second batch has $(D - \tau_d)$ hours to grow, so that at the final time they would total:

$$(Cell\ total\ at\ chemostat\ turnover\ time\ D)_{second\ batch} = V_d * X_0^{st} * e^{(D-\tau_d)/D}$$

In that manner, adding all the batches together we get:

$$(Cell\ total\ at\ chemostat\ turnover\ time\ D)_{all\ batches} = \cdots$$

$$\cdots = \sum_{n=0}^{n=\frac{D}{\tau_d}} V_d * X_0^{st} * e^{\frac{D-n*\tau_d}{D}} = \frac{D * V_d}{\tau_d} * X_0^{st} \sum_{n=0}^{n=D/\tau_d} e^{(D-n*\tau_d)/D}$$

Substituting $\frac{D*V_d}{\tau_d} = V_c$, and taking the equation to the limit we get:

$$\cdots = V_c * X_0^{st} \int_0^1 e^{(1-y)}\ dy = V_c * X_0^{st} * (e - 1)$$

So the total amount of cells outside the chemostat is $V_c * X_0^{st} * (e - 1)$, and the total amount of cells within the chemostat is $V_c * X_0^{st}$, totaling $e * V_c * X_0^{st}$.

Finally, the growth time required to reach $e * V_c * X_0^{st}$ cells, from an initial amount of $V_c * X_0^{st}$ cells, and a growth rate of D, is of course, 1/D hours.

# B) Basic "steady state" growth

### B5) What is the average cell growth rate at steady state?

<u>The growth rates of both cell strains, $\boldsymbol{\mu_0, \mu_1}$, through their Monod curves</u>

As mentioned above, we shall assume that the momentary growth rate of the reference strain follows Michaelis-Menten growth rate dynamics regarding a limiting substrate (i.e. Monod curve).

$$\mu_0(D) = \mu_0^{max} \cdot \left( \frac{S_0^{st}(D)}{S_0^{st}(D) + K_0} \right)$$

Also, we shall assume that the momentary growth rate of the mutant strain follows dynamics similar to Monod, with a potential "baseline growth rate" (growth with no available substrate), as follows:

$$\mu_1(D) = (\mu_1^{max} - \mu_1^{min}) \cdot \left( \frac{S_0^{st}(D)}{S_0^{st}(D) + K_1} \right) + \mu_1^{min}$$

Since, as we saw, $s_0$ is a function of D, the momentary growth rates of the strains, $\mu_i$, are also a function of D. This supports the claim that D is the main experimentally-controllable parameter.

Calculating the momentary growth rate for the substrate in the chemostat, $S_0^{st} = \frac{KD}{(\mu_{max} - D)}$:

$$\mu_0(D) = \mu_0^{max} \cdot \left( \frac{\frac{K_0 D}{(\mu_0^{max} - D)}}{\frac{K_0 D}{(\mu_0^{max} - D)} + K_0} \right) = \mu_0^{max} \cdot \left( \frac{K_0 D}{K_0 D + K_0 \mu_0^{max} - K_0 D} \right) = D$$

As expected.

And for the mutant strain:

$$\mu_1(D) = (\mu_1^{max} - \mu_1^{min}) \cdot \left( \frac{\frac{K_0 D}{(\mu_0^{max} - D)}}{\frac{K_0 D}{(\mu_0^{max} - D)} + K_1} \right) + \mu_1^{min} = \frac{\mu_1^{max} D + \mu_1^{min} \frac{K_1}{K_0} (\mu_0^{max} - D)}{D + \frac{K_1}{K_0} (\mu_0^{max} - D)}$$

# C) Mutation

In this chapter we estimate the time and the number of chemostat doublings needed, on average, for a desired mutated strain to appear. It will be shown that in some cases the desired mutations should appear quickly, and in other cases they are highly unlikely as follows:

A desired mutation of one specific SNP should appear after only few chemostat doublings, for a sufficiently large chemostat ($10^{11}$ cells or above). Also, a strain with two desired SNPs, where only their combination gives a fitness advantage whereas each one separately is neutral, is likely to appear only if the target size (the number of SNP locations that give rise to an advantageous mutation) for each SNP is very large. In other cases (e.g. two SNPs with a small target size, more SNPs or smaller chemostats) the appearance of the desired strain is highly unlikely, and is expected to require so much time to make the attempt ineffective unless each mutation has a benefit by itself.

Additionally, it will be shown that for a large chemostat we expect the time until an advantageous mutation occurs ($T_{mutant}$) to be substantially shorter than the time for an advantageous strain to take over the chemostat population ($T_{takeover}$, as will be addressed in section D), and not necessarily so in a small chemostat.

Note that this chapter discusses the mutation process only. The accumulation of mutations and genetic variation through successive advantageous mutations that take over and become fixed in the population is discussed in the following chapters. To illustrate these general principles, the questions will be broken up into a few steps, as follows:

(C1) How many cell doublings, on average, are needed for a specific desired SNP to appear? (C2) How many chemostat doublings, on average, are needed for a specific desired SNP to appear? How is this dependent on the size of the chemostat population? (C3) How many cell doublings and chemostat doublings, on average, are needed for a strain with two specific SNPs to appear? (C4) How many cell doublings and chemostat doublings are needed when the target size (θ) for a desired advantageous mutation is larger than one? (C5) How many cell doublings and chemostat doublings are needed in the general case for N SNPs and a target size θ?

After that, a few special cases are addressed:

(C6) How many cell and chemostat doublings, on average, are needed for every single SNP to appear? (C7) How do the results vary with different species? (C8) What is the effect of a hyper-mutating strain?

**C1) How many cell doublings, on average, are needed for a specific desired SNP to appear?**

Defining the size of the genome of the species to be $G = genome\ size\ [bp]$, and defining the mutation rate per base pair per replication to be $R_{mut} = mutation\ rate\ \left[\frac{SNP}{bp*doubling}\right]$, we get that the number of expected SNPs per cell doubling is:

$$P_{SNP} = G \cdot R_{mut} \left[\frac{SNP}{cell * doubling}\right]$$

For our *E. coli* test case $G = 5 \cdot 10^6\ [bp]$ [using BNID 103246]. Using $R_{mut} = 0.2 - 3 \cdot 10^{-10} \left[\frac{SNP}{bp*doubling}\right]$ [using the range found in BNID 100263, 111230, 111339, 111340], we get:

$$P_{SNP} = G \cdot R_{mut} \approx 10^{-4} - 10^{-3} \left[\frac{SNP}{cell * doubling}\right]$$

Or in other words, for every $10^3 - 10^4$ *E. coli* divisions we expect a single SNP mutation. This result is in concurrence with values from the literature [BNID 111339, 104475, 106748]. For simplicity, we shall use the conservative value $R_{mut} = 10^{-10} \left[\frac{SNP}{bp*doubling}\right]$.

Cell doublings for specific SNP

The probability of a specific SNP that represents the "correct" mutation appearing during a cell replication is:

$$p_{(SNP\ in\ correct\ position)} = (probability\ of\ SNP) \cdot (probability\ the\ SNP\ is\ in\ right\ place)$$

$$p_{(SNP\ in\ correct\ position)} = P_{SNP} \cdot \frac{1}{G} = R_{mut} = 10^{-10}$$

And for simplicity we will denote $p_{(SNP\ in\ correct\ position)}$ as $p$.

The number of trials needed on average to achieve at least one cell with the desired SNP equals $\frac{1}{p}$. Hence, the number of individual cell doublings we need to get the desired mutant strain is approximately:

$$doublings_{cell} \approx \frac{G}{P_{SNP}} = \frac{1}{R_{mut}} = 10^{10}\ [doublings]$$

(Mathematically, this is because the probability is distributed geometrically. A Bernoulli trial describes a trial that has only two possible outcomes, failure and success, where the probability of success is $p$. A geometric distribution describes the probability distribution of the number of sequenced Bernoulli trials needed to get one success. In our case each SNP represents a trial, and a SNP in the correct position represents "success". The number of trials needed on average, or with a probability of 50%, to achieve at least one success equals the mean of the distribution, which is $\frac{1}{p}$).

## C2) How many chemostat doublings, on average, are needed for a specific desired SNP to appear? How is does this depend on the size of the chemostat population?

The transition from cell doublings to chemostat doublings is conceptually very simple but we perform it in detail to show the major impact of the chemostat size and density. The number of chemostat doublings required depends on the number of single cell doublings required and the number of cells in the chemostat. The number of single cell doublings is analyzed above (section C1).

The number of cells in a standard chemostat is usually between $10^7$ cells for a "small and dilute" chemostat ($V_C = 0.01L, OD_{600} = 0.001\ E.coli,\ OD_{600} = 0.004\ Budding\ Yeast$) and $10^{12}$ cells for a "large" chemostat ($V_C = 1L, OD_{600} = 1\ E.coli,\ OD_{600} = 4\ Budding\ Yeast$) (see section A4 above). For our default characteristic values $N_{cells} = 10^{12} \cdot V_C \cdot OD_0 = 3 \cdot 10^{11}$.

The number of chemostat doublings needed to achieve $doublings_{cell}$, and therefore the desired SNP, is:

$$doublings_{chemostat} = \frac{doublings_{cell}}{N_{cells}} = \frac{1}{R_{mut} \cdot N_{cells}}$$

and for *E. coli*:

$$doublings_{chemostat} = \frac{1}{R_{mut}} \cdot \frac{1}{OD_0 \cdot V_C \cdot 10^{12}}$$

Using our default characteristic numbers:

$$doublings_{chemostat} = \frac{10^{10}}{3 \cdot 10^{11}} = 0.03$$

Therefore we can expect a mutant strain with one desired SNP to be present in the chemostat after less than one doubling in our default characteristic case, and certainly in the "large" $10^{12}$ cell chemostat.

For a "small" chemostat of $10^7$ cells the number of chemostat doublings needed is:

$$doublings_{chemostat} = \frac{10^{10}}{10^7} = 1000\ doublings$$

For a doubling time of approx. 10 hours, we get an impractical amount of time until the one-SNP mutation appears on average, or:

$$Total\ Time \approx 10,000\ hours \approx 1\ year$$

## C3) How many cell doublings and chemostat doublings, on average, are needed for a strain with two specific SNPs to appear?

Single cell doublings

It can be assumed that the first and second SNP in each strain occur independently, so the number of SNPs required is squared, therefore:

$$p_{(2\ SNPs\ in\ correct\ position)} = (prob.\ of\ 2\ SNPs) \cdot (prob.\ 2\ SNPs\ are\ in\ place)$$

$$p_{(2\ SNPs\ in\ correct\ position)} = (prob.\ of\ 1\ SNP)^2 \cdot \frac{1}{2}(prob.\ 1\ SNP\ is\ in\ place)^2$$

$$p_{(2\ SNPs\ in\ correct\ position)} = \frac{1}{2}p_{(1\ SNP\ in\ correct\ position)}^2$$

where the factor of 1/2 comes from the fact that the order does not matter between the first and second SNP.

$$doublings_{cell} \approx \frac{\frac{1}{2}G^2}{P_{SNP}^2} = \frac{1}{2} \cdot \frac{1}{R_{mut}^2}$$

For our default characteristic values

$$doublings_{cell} \approx 5 \cdot 10^{19}\ [doublings]$$

Chemostat doublings

As before, the transition from cell doublings to chemostat doublings depends on the number of single cell doublings required and the number of cells in the chemostat.

$$doublings_{chemostat} = \frac{doublings_{cell}\ for\ 2\ SNPs}{N_{cells}}$$

$$doublings_{chemostat} = \frac{1}{2} \cdot \frac{1}{R_{mut}^2 \cdot N_{cells}}$$

Using our default characteristic numbers, $doublings_{chemostat} \approx 10^{10}$, meaning it should take over a million years for two specific SNPs to appear in the chemostat.

To understand the vast difference between one and two SNPs, note that the first SNP has a background strain of $N_{cells}$ from which to evolve, but the second SNP in the combination only has the cells that accumulated the first mutation in the first step. This means that the cell doublings required for the second SNP do not occur via many separate cells in parallel as for the first SNP, but via a small number of cells doubling repeatedly in succession.

In conclusion, a mutant strain that requires two (or more) specific SNPs together to gain a selective advantage is very unlikely in a chemostat setting. This finding may seem counterintuitive, since in practice mutants appear with more than one SNP. There are several ways a mutant strain with two (or more) SNPs can take over the chemostat population: when only one SNP gives the advantage (the "driver") and the other SNP is random ("passenger"), when each SNP already gives a selective advantage on its own and successive takeover sweeps take place (as will be shown in chapter E), and when the target size for each SNP is sufficiently large (as will be analyzed in the following questions).

**C4) How many cell doublings and chemostat doublings are needed when the target size ($\theta$) for a desired advantageous mutation is larger than one?**

Cell doublings

The number of cell doublings required is calculated similarly to the case of a specific desired SNP (section C1 above), but with a higher probability of success $p$. Defining target size $\theta$ to be the number of SNP locations that give rise to an advantageous mutation, the probability of success, or a SNP within the desired target, is:

$$p_{(SNP\ in\ target)} = \frac{target\ size}{genome\ size} * probability\ of\ SNP$$

$$p_{(SNP\ in\ target)} = \frac{\theta}{G} \cdot P_{SNP} = \theta \cdot R_{mut}$$

The number of cell and chemostat doublings needed to achieve an advantageous mutant strain is:

$$doublings_{cell} = \frac{1}{p} = \frac{1}{\theta \cdot R_{mut}}\ [doublings]$$

$$doublings_{chemo\prime} = \frac{1}{\theta \cdot R_{mut} \cdot N_{cells}}\ [doublings]$$

For our default characteristic values and for a target size of $\theta = 1000\ [bp]$ (target sizes may vary, we chose a value on the order of the length of a gene, similar to that used in [2] whose experiment resembles our setup, and similar to [43], [44]):

$$doublings_{cell} = 10^7\ [doublings]$$

$$doublings_{chemostat} \approx \begin{cases} 10^{-5} for\ a\ large\ chemostat\ (10^{12} cells, OD_{600} = 1, V_C = 1L) \\ 10^1\ small\ chemostat\ (10^6 cells, OD_{600} = 0.001, V_C = 0.001L) \end{cases}$$

We reach an important conclusion here. In practice, for a "large" chemostat we expect the time until an advantageous mutation occurs ($T_{mutant}$) to be substantially shorter than the time for an advantageous strain to take over the chemostat population ($T_{takeover}$, as will be addressed in section D), and not necessarily so in a "small" chemostat.

This difference between "large" and "small" chemostats was experimentally analyzed in the work of T. Egli et. al. [2], where a chemostat of $10^{11}$ cells showed "clock-like" repetitive takeover sweeps and a chemostat of $10^7$ cells showed only random takeover sweeps with long intervals between them. These results also fit with the analysis of [45].

**C5) How many cell doublings and chemostat doublings are needed in the general case for N SNPs and a target size $\theta$?**

Putting together the results from the above questions, the probability of N SNPs in N independent targets of target size $\theta$ is:

$$p_{(N\ SNPs\ in\ N\ targets)} = \frac{1}{N!}\left(p_{(SNP\ in\ target)}\right)^N = \frac{1}{N!}(\theta \cdot R_{mut})^N$$

Where the factor of $\frac{1}{N!}$ comes from the fact that the order of the mutations is not important.

And the number of cell and chemostat doublings required is:

$$doublings_{cell} = \frac{1}{N!} \cdot \frac{1}{(\theta \cdot R_{mut})^N}\ [doublings]$$

$$doublings_{chemostat} = \frac{1}{N!} \cdot \frac{1}{(\theta \cdot R_{mut})^N \cdot N_{cells}}\ [doublings]$$

In an order of magnitude calculation, for a mutation rate of $R_{mut} \approx 10^r$, a target size of $\theta \approx 10^g$ and a chemostat with a volume and OD corresponding to $N_{cells} \approx 10^n$ we get $doublings_{chemostat} \approx 10^{-Ng-Nr-n}$. For the desired mutation to appear quickly we require the number of chemostat doublings needed on average to be small, or:

$$(-Ng - Nr - n) \leq 0$$

and for r = 10, n = 11 we require the target size to be:

$$g \geq 10 - \frac{11}{N}$$

For N=1 SNP any target size is sufficient. For N=2 SNPs the required target size is on the order of $10^{4-5}$, which is possible, but not common. N>2 SNPs are impossible to achieve in a single doubling of a normal sized chemostat, regardless of the target size. (When the target size reaches the size of the genome there are no longer multiple independent targets, and the assumptions of the calculation above do not hold).

In summary, for a sufficiently large chemostat, an advantageous mutant strain with a desired SNP is expected to appear in a short period of time. A strain that needs two SNPs together to become advantageous is possible only when the target sizes are sufficiently large. Other mutations are expected only through successive advantageous takeover sweeps, but not if they are all needed to exist concurrently to give any fitness advantage directly. These results show that successive takeovers are the only reliable way (i.e. that does not rely on luck against impossibly low odds) for evolution to proceed in a chemostat.

**C6) How many cell and chemostat doublings, on average, are needed for every single SNP to appear?**

In this question we refer to a seemingly extreme scenario where we require every possible single SNP to exist in at least one cell in the chemostat. It is shown that this scenario is actually quite likely, for a sufficiently large chemostat. On the other hand, as was shown above (section C3) a combination of two specific SNPs is almost impossible (let alone every combination of two SNPs). These results can help give intuition to the genetic variability expected inside a chemostat.

A SNP in every possible position within the population:

There are G positions in which a SNP can occur. Clearly, only one SNP is needed for the first strain to have a SNP in a unique position. Next, there are (G-1) positions where a new SNP will have a unique position (i.e. different from the first existing mutant strain) out of the possible G positions. Therefore, for a second strain to appear with a SNP in a new unique position $\frac{G}{G-1}$ SNPs are needed, on average. For a third unique strain $\frac{G}{G-2}$ SNPs are needed on average, and so forth. Accordingly, the number of SNPs needed for there to be a SNP in every possible position is:

$$number\ of\ SNPs\ needed = \left(1 + \frac{G}{G-1} + \frac{G}{G-2} + \cdots + \frac{G}{1}\right) = G \cdot \sum_{j=1}^{G} \frac{1}{j} \approx G \cdot \ln G$$

The number of single cell doublings needed for there to be a SNP in every possible position is:

$$doublings_{cell} = \frac{number\ of\ SNPs\ needed}{probability\ of\ each\ SNP}$$

$$doublings_{cell} \approx \frac{G \cdot \ln G}{P_{SNP}}$$

Using $P_{SNP} = G \cdot R_{mut}$ from above:

$$doublings_{cell} \approx \frac{\ln G}{R_{mut}}$$

Using our default characteristic numbers:

$$doublings_{cell} \approx 10^{11}[doublings]$$

Chemostat doublings

As in the cases in the above sections (C2-C5) the number of chemostat doublings needed to achieve the required $doublings_{cell}$, and therefore every single SNP, is:

$$doublings_{chemostat} = \frac{doublings_{cell}}{N_{cells}} = \frac{\ln G}{R_{mut} N_{cells}}$$

and for *E. coli*:

$$doublings_{chemostat} = \frac{\ln G}{R_{mut}} \cdot \frac{1}{OD_0 \cdot V_C \cdot 10^{12}}$$

Using our default characteristic numbers:

$$doublings_{chemostat} = \frac{10^{11}}{2.5 \cdot 10^{11}} = 0.4$$

Meaning, we can expect every possible one SNP mutant to be present in the chemostat after less than one doubling in our default characteristic case, and certainly in the "large" $10^{12}$ cell chemostat.

For a "small" chemostat of $10^7$ cells the number of chemostat doublings needed is:

$$doublings_{chemostat} = \frac{10^{11}}{10^7} = 10,000 \; doublings$$

For a doubling time of approx. 10 hours, we get: $Total\;Time \approx 100,000\;hours \approx 10\;years$.

Comparing to the required number of doublings required for every single SNP to the number of doublings required for a specific SNP (section C1-C2), we find:

$$doublings_{cell}\;for\;every\;SNP = \ln G \cdot (doublings_{cell}\;for\;specific\;SNP)$$
$$doublings_{chemostat}\;for\;every\;SNP = \ln G \cdot (doublings_{chemostat}\;for\;specific\;SNP)$$

In summary, for a sufficiently large chemostat every possible single SNP is expected to appear, but not every combination of two SNPs. Therefore, strains with more than one SNP are substantially more likely to exist if the first SNP has a selective advantage.

# D) Single takeover

**D2) How does an extended version of the Monod relationship change the takeover time and dynamics?**
For the completeness of the analysis presented here, we note the results if we were to work with an extended Monod model as given by Korvarova [21], [20] (developed to take in to account a non-zero minimal limiting substrate concentration below which there is no growth):

$$\mu_0 = \mu_0^{max} \cdot \left( \frac{S_0^{st} - S_{min}}{S_0^{st} + K_0 - S_{min}} \right)$$

and the respective:

$$\mu_1(D) = (\mu_1^{max} - \mu_1^{min}) \cdot \left( \frac{S_0^{st} - S_{min}}{S_0^{st} + K_1 - S_{min}} \right) + \mu_1^{min}$$

We solve by working backwards, and requiring the steady state growth rate to be D:

$$\mu_0 = \mu_0^{max} \cdot \left( \frac{S_0^{st} - S_{min}}{S_0^{st} + K_0 - S_{min}} \right) \stackrel{!}{=} D$$

$$D \cdot S_0^{st} + D \cdot K_0 - D \cdot S_{min} = \mu_0^{max} \cdot S_0^{st} - \mu_0^{max} S_{min}$$

$$S_0^{st}(\mu_0^{max} - D) = S_{min}(\mu_0^{max} - D) + D \cdot K_0$$

$$\boxed{S_0^{st} = S_{min} + \frac{K_0 D}{(\mu_0^{max} - D)}}$$

$$\mu_1(D) = (\mu_1^{max} - \mu_1^{min}) \cdot \left( \frac{S_{min} + \frac{K_0 D}{(\mu_0^{max} - D)} - S_{min}}{S_{min} + \frac{K_0 D}{(\mu_0^{max} - D)} + K_1 - S_{min}} \right) + \mu_1^{min} =$$

$$\dots = (\mu_1^{max} - \mu_1^{min}) \cdot \left( \frac{\frac{K_0 D}{(\mu_0^{max} - D)}}{\frac{K_0 D}{(\mu_0^{max} - D)} + K_1} \right) + \mu_1^{min} = back\ to\ the\ original\ problem$$

So that:

$$\mu_0(D)_{Monod} = \mu_0(D)_{Kovarova}$$
$$\mu_1(D)_{Monod} = \mu_1(D)_{Kovarova}$$

Since we are not interested in the absolute substrate values, but rather the growth rates determined by the substrate levels determined by the dilution rate D, for our purposes there is no difference between the models and we shall use the basic Monod model.

**D3) How long does it take from the first appearance of a mutant until takeover?**

**The takeover time** $T_{takeover} = \frac{\ln(\text{number of doublings})}{\Delta\mu}$

The time of cellular growth is:

$$X = X_{(t=0)}e^{\mu_{eff}t} \rightarrow t = \frac{\ln\left(\frac{X}{X_{(t=0)}}\right)}{\mu_{eff}}$$

$$growth\ time = \frac{\ln\left(\frac{final\ number\ of\ cells}{inital\ number\ of\ cells}\right)}{effective\ growth\ rate}$$

In our case we get:

$$T_{takeover} = \frac{\ln\left(\frac{N_{cells}*10\%}{1}\right)}{\Delta\mu}$$

Where the initial amount of cells is one mutant cell and $N_{cells}*10\%$ (10% of the cells in the chemostat) is the amount of cells the mutant strain reach to have taken over the population (in section A4 the value of $N_{cells}$ is calculated).

And then the takeover time is:

$$T_{takeover} = \ln(N_{cells}) \cdot \frac{D + \frac{K_1}{K_0}(\mu_1^{max} - D)}{D^2\left[\frac{K_1 - K_0}{K_0}\right] + D\left[\mu_1^{max} + \frac{K_1}{K_0}(\mu_1^{min} - \mu_0^{max})\right] + \left[\frac{K_1}{K_0} \cdot \mu_1^{min} \cdot \mu_0^{max}\right]}$$

We see that the main parameter that is experimentally controllable is the chemostat dilution parameter D, which has no cost to change, as well as $S_d$ and $V_t$.

## E) Successive takeover

To answer the questions in this chapter, we created a time-step dynamic computer simulation. In this simulation, multiple strains take over one another, where each new strain is supposed to have mutated out of its predecessor. Strain i=1 "mutates" out of strain i=0, strain 1=2 from i=1 etc. The stages of the simulation are:

(1) **SETUP:** The starting conditions are:
   - A chemostat with cell strain i=0 at steady state ($X_0^{initial} = X_0^{st}$, $S^{initial} = S_0^{st}$).
   - All other strains don't yet exist $X_{i>0} = 0 \left[\frac{g}{L}\right]$.

(2) **DILUTION:** A time step starts at the end of the previous step with proper influx and outflux:
   - Dilution of the cell strains in the chemostat: $X_i(new) = \phi \cdot X_i(old)$
   - Dilution of the limiting substrate in the chemostat, and influx of a new drop of limiting substrate. $S(new) = \phi \cdot S(old) + (1-\phi) \cdot S_d$.

(3) **GROWTH:** Each time step then proceeds with growth for a time of $\tau_d$, until the next dilution drop is to occur. The growth is determined dynamically by the Monod curves of the strains, for the current limiting substrate concentration. In other words, a set of coupled differential equations (one for each one of the i strains, and another for the limiting substrate) is solved using a numerical ODE solver, with starting conditions set by the dilution of the previous step.

(4) **MUTATION**: there are two parameters that define all the new mutant strains – what the "parent" strains population needs to be for the child to appear (representing the target size $\theta$), and the improvement in affinity to the limiting substrate (For simplicity, we assumed that both of these parameters stay constant throughout the simulation, i.e. $\lambda \equiv \frac{k_{i+1}}{k_i} = \frac{k_{i+2}}{k_{i+1}}$, etc.):
   - When a cell strain populations reaches a predetermined size, $X_i >$ the concentration of $10^n$ cells than a new strain emerges $X_{i+1} =$ concentration of a single cell
   - The improvement in affinity to the limiting substrate: $\lambda \equiv \frac{k_{i+1}}{k_i}$

(5) **FINAL TAKEOVER:** The "final" takeover represents the new strain that can grow utilizing other substrates and not only the limiting substrate. This is represented by a higher affinity and a higher yield (the yield, in that more biomass can be produced for the same amount of limiting substrate; The other mutant improvements only increase the **rate** in which biomass is produced, but not the **total amount**, because of mass conservation). See section B2 for further understanding.

## Acknowledgements

The authors would like to thank Niv Antonovsky, Ghil Jona, Ron Sender, Dan Davidi, Yinon Moise Bar-On for their help and advice in this study. A special thanks to Avi Flamholtz, Shmuel Gleizer and Noam Prywes for their insights and review of the work in progress and to Elad Noor for his helpful comments and help with the mathematical framework.